\documentclass[journal]{IEEEtran}
\usepackage{setspace}

\usepackage{changes}
\definechangesauthor[name=Reviewer 2, color = red]{r2}
\definechangesauthor[name=Reviewers 3, color = orange]{r3}
\definechangesauthor[name=improvement, color = blue]{imp}
\ifCLASSINFOpdf
\else
\fi
\usepackage[cmex10]{amsmath}

\usepackage{caption}
\usepackage{subcaption}

 \usepackage[hyperindex=true,pdftitle={},
pdfauthor={St\'ephane Guerrier},colorlinks=TRUE,
pagebackref=false,citecolor=blue,plainpages=false,
pdfpagelabels]{hyperref} 

\usepackage{verbatim} 
\usepackage{amsthm}
\usepackage{amsbsy}
\usepackage{amssymb}
\usepackage{amsbsy}
\usepackage{amsfonts}
\usepackage{graphics}
\usepackage{bm}
\usepackage{bbm}
\usepackage{mathtools}
\usepackage{authblk}
\usepackage{booktabs}
\usepackage{nicefrac}
\usepackage{enumitem}
\usepackage{bbding}
\usepackage{soul}
\usepackage[numbers]{natbib}
\usepackage{nicefrac}
\usepackage{placeins}
\usepackage{xcolor}
\usepackage{siunitx}
\usepackage{tabularx,booktabs}
\usepackage{algorithm,algorithmic}

\usepackage{hyperref} 

\DeclareMathOperator*{\argmin}{argmin}

\DeclareMathOperator*{\var}{var}

\DeclareSymbolFont{lettersA}{U}{txmia}{m}{it}
\DeclareMathSymbol{\real}{\mathord}{lettersA}{"92}
\DeclareMathSymbol{\field}{\mathord}{lettersA}{"83}

\hypersetup{pdfauthor={Name}}

\DeclarePairedDelimiter\floor{\lfloor}{\rfloor} 

\newtheoremstyle{mytheoremstyle} 
    {0.3cm}                      
    {0cm}                        
    {\itshape}                   
    {}                           
    {\scshape}                   
    {: }                          
    {0em}                       
    {}  

\theoremstyle{mytheoremstyle}
\newtheorem{Theorem}{Theorem}

\newtheorem{Corollary}{Corollary}

\renewenvironment{proof}{{\noindent \sc Proof:}}{\qed}

\newtheoremstyle{myExampleRemarkstyle} 
    {0.3cm}                    
    {0cm}                           
    {\itshape}                   
    {}                           
    {\scshape}                   
    {: }                          
    {0em}                       
    {}  

\theoremstyle{myExampleRemarkstyle}
 
\newtheorem{Remark}{Remark}

\newtheoremstyle{simuStyle}
{0.3cm} 
{0cm} 
{} 
{} 
{\bfseries} 
{.} 
{0em} 
{} 

\theoremstyle{simuStyle}

\newtheoremstyle{stratStyle}
{0.3cm} 
{0cm} 
{} 
{} 
{\scshape} 
{: } 
{0em} 
{} 

\theoremstyle{stratStyle}

\DeclareSymbolFont{lettersA}{U}{txmia}{m}{it}
\DeclareMathSymbol{\real}{\mathord}{lettersA}{"92}
\DeclareMathSymbol{\field}{\mathord}{lettersA}{"83}
\def\simiid{\stackrel{iid}{\sim}}

\def\real{{\rm I\!R}}

\def\0{{\bf 0}}

\def\bOmega{{\bm{\Omega}}}

\def\X{{\bf X}}

\def\I{{\bf I}}
\def\Y{{\bf Y}}

\def\A{{\bf A}}

\def\V{{\bf V}}
\def\c{{\bf c}}

\def\b{{\bf b}}

\def\G{{\bf G}}
\def\g{{\bf g}}
\def\W{{\bf W}}
\def\bomega{{\boldsymbol \omega}}
\def\bg{{\boldsymbol \gamma}}
\def\bSigma{{\boldsymbol \Sigma}}
\def\1{{\boldsymbol 1}}

\def\trans{^{\top}}

\def\0{\mathbf{0}}

\def\wh{\widehat}
\def\wt{\widetilde}

\newcolumntype{Y}{>{\centering\arraybackslash}X}

\def\boxit#1{\vbox{\hrule\hbox{\vrule\kern2pt  \vbox{\kern2pt#1\kern2pt}\kern2pt\vrule}\hrule}}

\begin{document}
%
\title{Scale-wise Variance Minimization for \\ Optimal Virtual Signals: \\ An Approach for Redundant Gyroscopes}
%
%
%

\author{Yuming~Zhang, Davide~A.~Cucci, Roberto~Molinari, St\'ephane~Guerrier
\thanks{This work was supported in part by the SNSF Professorships Grant 176843, in part by the Innosuisse-Boomerang Grant 37308.1 IP-ENG.}
\thanks{\textbf{Y. Zhang} is PhD candidate, Geneva School of Economics and Management, University of Geneva, 1205, Switzerland. (E-mail: \texttt{Yuming.Zhang@unige.ch}).}
\thanks{\textbf{D. Cucci} is Senior Research Associate, Geneva School of Economics and Management, University of Geneva, 1205, Switzerland. (E-mail: \texttt{Davide.Cucci@unige.ch}).} 
\thanks{\textbf{R. Molinari} is Assistant Professor, Department of Mathematics and Statistics, Auburn University, AL 36849, USA. (E-mail: \texttt{robmolinari@auburn.edu})}
\thanks{\textbf{S. Guerrier} is Assistant Professor, Faculty of Science \& Geneva School of Economics and Management, University of Geneva, 1205, Switzerland. (E-mail: \texttt{Stephane.Guerrier@unige.ch}).}}%

\maketitle

\begin{abstract}
The increased use of low-cost gyroscopes within inertial sensors for navigation purposes, among others, has brought to the development of a considerable amount of research in improving their measurement precision. Aside from developing methods that allow to model and account for the deterministic and stochastic components that contribute to the measurement errors of these devices, an approach that has been put forward in recent years is to make use of arrays of such sensors in order to combine their measurements thereby reducing the impact of individual sensor noise. Nevertheless combining these measurements is not straightforward given the complex stochastic nature of these errors and, although some solutions have been suggested, these are limited to certain specific settings which do not allow to achieve solutions in more general and common circumstances. Hence, in this work we put forward a non-parametric method that makes use of the wavelet cross-covariance at different scales to combine the measurements coming from an array of gyroscopes in order to deliver an optimal measurement signal without needing any assumption on the processes underlying the individual error signals. As a result of this work we also study an appropriate non-parametric approach for the estimation of the asymptotic covariance matrix of the wavelet cross-covariance estimator which has important applications beyond the scope of this work. The theoretical properties of the proposed approach are studied and are supported by simulations and real applications, indicating that this method represents an appropriate and general tool for the construction of optimal virtual signals that are particularly relevant for arrays of gyroscopes. Moreover, the results of this work can support the creation of optimal signals for other types of inertial sensors other than gyroscopes as well as for redundant measurements in other domains other than navigation.
\end{abstract}

\begin{IEEEkeywords}
Virtual Gyroscopes, Wavelet Variance, Redundant Sensor Arrays, Inertial Sensor Calibration, Sensor Fusion.
\end{IEEEkeywords}

\IEEEpeerreviewmaketitle

\section{Introduction}
\label{sec.intro}


Inertial sensors, such as accelerometers and gyroscopes, are ubiquitous in modern navigation systems, with applications in satellites, aviation, drones, and personal navigation in smartphones. They provide high-frequency and short-term precise information on the change in the orientation and velocity that can be used to reconstruct the motion of objects and vehicles they are placed on. Similarly to others, these sensors are characterized by errors that can be both deterministic and stochastic. Deterministic errors, such as the stable parts of scale factors and axes non-orthogonality, can be pre-calibrated and removed from the measurements directly~\cite{clausen2019calibration}. The stochastic part of the error, such as turn-on biases, bias instability and drift 
can only be corrected during navigation, up to a certain extent, provided that a suitable stochastic model has been determined~\cite{guerrier2013wavelet, el2008analysis}. If inadequately taken account of, stochastic errors can lead to macroscopic drifts in the estimated position and orientation when inertial sensors are integrated alone over a sustained amount of time. Also as a consequence of this feature, these sensors are typically employed along with other sensors such as global navigation satellite systems in strap-down inertial navigation~\cite{titterton2004strapdown}, and cameras in visual-inertial systems~\cite{huang2019visual}. 

Given the considerable impact of stochastic errors on navigation performance, one of the directions that researchers and manufacturers have been focusing on is to create devices with better stochastic properties, such as lower white noise power spectral densities or better bias stability, which allow to classify inertial sensors in industrial, tactical or navigation grade. To date, the latter devices deliver excellent performance for the current application domains but tend to be bulky and expensive. As an alternative, the technological development in  Micro Electrical Mechanical System~(MEMS) has revolutionized the inertial sensor industry over the last decade~\cite{shaeffer2013mems}. Nowadays, six-axis Inertial Measurement Units~(IMUs), composed of three accelerometers and three gyroscopes, can be bought in considerable volumes at exceedingly reasonable costs and are widely employed as motion sensors in consumer electronics, such as smartphones, but also in drones and in industrial manufacturing equipment. However, despite the recent developments, currently available low-cost MEMS cannot fully meet the requirements of many applications, especially when long-term bias stability is required~\cite{perlmutter2016future}.

With the aim of addressing the above limitation of MEMS IMUs, increased research has been focused on taking advantage of their small size, cost and power consumption allowing the construction of arrays of such sensors, for instance placing multiple off-the-shelf sensor triads in a planar configuration, or single axis sensors in a non-planar one~\cite{song2019data}. Partially redundant observations of the same quantity can be fused to compute synthetic measurements with better stochastic properties compared to those of the single sensors. Additionally, the uncertainty of such measurements can be better quantified, their dynamic range can be extended, and a higher system robustness can be achieved through fault detection and isolation mechanisms~\cite{guerrier2012fault}. For an extensive literature review on inertial sensor arrays we refer the readers to~\cite{nilsson2016inertial}.

In the case of redundant gyroscopes, multiple observations of the platform angular velocities are readily available to be fused together~\cite{vaccaro2017reduced}, once small misalignment and non-orthogonality in sensor axes have been corrected~\cite{clausen2019calibration}. For accelerometers, this is slightly more complicated since the specific force measured by each sensor depends on the sensor placement and on the angular velocity. However, this allows to create \textit{gyro-free} IMUs that recover the angular velocity from specific force readings at different but known locations~\cite{skog2016inertial, cucci2016analysis}. Other authors have been focusing on the fusion of entire IMUs in different configurations~\cite{clausen2015position, waegli2010noise} and the approach proposed in this work can be employed in this direction. However, despite the achievements of current research on the optimal fusion of MEMS-IMUs (stochastic) measurements, these results are limited in scope since they only work under narrow parametric assumptions for the stochastic error which is also assumed to be identical for all gyroscopes in the array. For example, in~\cite{vaccaro2017reduced} the gyroscope stochastic errors are assumed to be composed only by a white noise (also known as angular random walk) and a random walk (also known as rate random walk) processes based on which they propose a method to determine the coefficients of the linear combination of individual gyroscope signals such that the variance of the random walk is minimized. While the latter is a good solution and is the only method currently available for this purpose, the underlying parametric assumptions may often not hold in practice, especially when low-cost MEMS gyroscopes are considered since they tend to have much more complex stochastic structures.

Considering the above and developing on the proposal in \cite{zhang2018optimal}, this work aims to formally define and study a broader \textit{non-parametric} method to build an optimal virtual signal which targets arrays of gyroscopes but also provides the bases to construct virtual signals for other applications.
{\color{black} To define optimality, we consider a quantity called Wavelet Variance~(WV), which is extremely informative for the stochastic properties of a signal over different time-scales \cite{percival2006wavelet}. When the signal is stationary, the WV provides an exact scale-wise decomposition of the signal's variance. The WV is also well-defined for signals that have non-stationary features such as random walk or drift, both of which are commonly observed in sensor stochastic errors. Therefore, the WV can provide a representation of the process variance even when these signals have infinite variance. As a consequence of these properties, it is possible to characterize (and therefore minimize) the process variance using this scale-wise representation.
Moreover, when using the Haar wavelet filter, the WV is equivalent (up to a constant) to the Allan variance, originally introduced to characterize the stability of atomic clocks~\cite{allan1966statistics} and now the standard for stochastic calibration of inertial sensors~\cite{el2008analysis}. The reason for the latter is that the behaviour of the WV at different time-scales can be related to the navigation performances of the resulting system~\cite{GMWM2014}. Therefore, we consider a linear combination of the sensor signals and we determine the coefficients of this linear combination such that the WV of the resulting virtual sensor is minimized at user specified scales.} In the latter direction, and depending on the application, the user would be able to freely choose which time-scales should be minimized by specifying a weight vector, and thus for example minimizing most of the variance at the high scales, approximately corresponding to bias-instability or rate-random-walk effects, or at the low scales, mostly corresponding to white noise, or at any combination of scales required.

With this goal in mind, this work is organized as follows. In Section~\ref{sec:syn:sensors}, we introduce the proposed method and we define an estimator for the coefficients of the linear combination of sensor signals (e.g. gyroscopes). In Section~\ref{sec:main:res}, we study the statistical properties of the proposed estimator. More precisely, we demonstrate the consistency and the asymptotic normality of our estimator. In addition, in Section~\ref{sec:comp} we study a non-parametric approach to compute the asymptotic covariance matrix of the proposed estimator, the results of which have important impacts beyond the scope of this work. Finally, in Sections~\ref{sec:study} and~\ref{sec:realcase} we present a comprehensive evaluation of our method based on Monte-Carlo simulations and real data, along with a comparison with the method presented in~\cite{vaccaro2017reduced}. Section~\ref{sec:conclusions} concludes.

\section{Optimal Combination of Gyroscope Signals}
\label{sec:syn:sensors}

{\color{black}As mentioned earlier, the proposed method can be broadly applied to any set of redundant signals which respect the conditions stated further on. Keeping this in mind, for simplicity in the rest of this work we will focus on gyroscope signals as they indeed fulfill these requirements.} To formally introduce the proposed approach, using the notation $\mathbb{N}_+ \vcentcolon= \mathbb{N} \setminus \{0\}$, let us define a multivariate process as $(\mathbf{X}_t: t=1,\ldots,T)$ with $T \in \mathbb{N}_+$ and $\mathbf{X}_t \in \real^p$ with $p \in \mathbb{N}_+$. Hereinafter, to simplify notation, we will refer to the multivariate process as $(\mathbf{X}_t)$ which in this setting represents the signals of the $p$ gyroscopes of length $T$ all measured along the same axis, but can also represent a general set of redundant measurement signals as mentioned in the introduction. Based on this, we also let $\X_t \in \real^p$ represent the vector of measurements from each gyroscope at time $t$ and let $(X_{i;t})$, with $i = 1, \ldots, p$, denote the signal of the $i$-th gyroscope. 

Without loss of generality, throughout this work we assume that calibration data is acquired in static conditions or under a  constant angular velocity $\delta$ (e.g. employing a rotation table). 
Thus, each gyroscope signal $(X_{i;t})$ corresponds to an unbiased measurement of the same constant velocity signal $\delta$, implying that $(X_{i;t})$ is such that $\mathbb{E}[X_{i;t}] = \delta$ for all $i = 1,\ldots,p$.

\begin{Remark}
\label{rem.deterministic}
In general, $\mathbb{E}[X_{i;t}] = \delta_t+b_i$, where $\delta_t$ denotes the angular velocity and $b_i$ denotes the per-gyroscope bias. The angular velocity $\delta_t$, aside from possibly depending on time, can be the result of the combined effects of non-orthogonality of sensor axes and scale factors, which should be corrected beforehand by means of deterministic calibration~\cite{clausen2019calibration}. On the other hand, the per-gyroscope bias $b_i$ can be handled either through individual detrending or directly through the time-differencing embedded in the method put forward in the following sections which, in addition, is solely based on the decomposition of the variance of $(X_{i;t})$.
\end{Remark}
\vspace{0.2cm}

Using the above notation and recalling \citep{zhang2018optimal}, we can now define a virtual gyroscope signal $(S_t)$, with $t=1,\hdots,T$, by taking a linear combination of the $p$ gyroscope signals as 
\begin{equation}
    S_t \vcentcolon= \c\trans \mathbf{X}_t,
    \label{eq:virtual:sensor}
\end{equation}
where $\mathbf{c} = (c_1, \ldots, c_p)\trans \in \boldsymbol{\mathcal{C}}$ corresponds to the vector of coefficients given to each signal with 
\begin{equation*}
    \boldsymbol{\mathcal{C}} \vcentcolon= \left\{\c\in\real^p: \c\trans\1_p=1\right\},
\end{equation*}
and $\1_p$ as a vector of length $p$ with all elements equal to one.

The constraint that the coefficients add up to one allows the virtual gyroscope to have the same constant velocity signal $\delta$ as the individual ones: 
\begin{equation*}
    \mathbb{E} [S_t] =  \mathbb{E} \left[ \sum_{i = 1}^p {c}_i X_{i;t}\right] = \delta \sum_{i = 1}^p {c}_i = \delta.
\end{equation*}
With the above framework, the goal of this work is therefore to define the optimal values for the coefficients $\c$. Since the expectation of this virtual signal remains unchanged under the above constraints, the most intuitive criterion to define these coefficients would therefore depend on the variance of the resulting virtual signal. Indeed, we would require the resulting virtual signal to be as ``precise'' as possible and hence to have the smallest measurement uncertainty among all possible virtual signals defined as above. This implies that the coefficients $\c$ should be chosen to ensure that the constructed virtual gyroscope has the smallest variance among all possible values of $\c$. 

While approaches to minimize the variance of linear combinations of random variables (signals) are well known in domains such as finance, due to the presence of processes that are commonly observed within the stochastic error of inertial sensors, the actual variance of $(S_t)$ may often be infinite. For example, if drift or random walk processes are present in $(\mathbf{X}_t)$, then the resulting virtual signal $(S_t)$ will also have a variance that explodes with the size of the signals (i.e. infinite variance) and will consequently not deliver coefficients that are consistently optimal over time. Additionally, the constructed virtual signal will be characterized by unknown statistical properties. To address this problem, the previously discussed WV (and related quantities) can be employed since, due to the inherent differencing properties of the wavelet decomposition, the WV exists and is finite also for various non-stationary processes with infinite variance. As a result, the wavelet representation of the process variance can also be used to characterize $(S_t)$ and consequently provide a criterion that can be minimized in a statistically meaningful manner.

Let us therefore formalize the definition of the WV to introduce the approach we put forward in this work. Let $J>0$ be a fixed integer that represents the last level of wavelet decomposition considered for each signal. The wavelet coefficients of the virtual gyroscope for level $j$, with $j=1,\ldots,J$, are defined as:
\begin{equation*}
    \wt{W}_{j,t} \vcentcolon = \sum_{l = 0}^{L_{j} - 1} h_{j, l} S_{t-l},
\end{equation*}
where $h_{j,l}$ is the $j$-th level wavelet filter with length $L_j\in\mathbb{N}_+$ and $l=0,\ldots,L_j-1$. For example, if we consider the Haar wavelet filter (one of the most commonly used filters in practice) we directly have $L_j = 2^j$. Given this, the WV at level $j$ is defined as the variance of the wavelet coefficients and is denoted as $\var(\wt{W}_{j,t})$. As highlighted earlier, this quantity is useful when there are non-stationary processes included in the signals and, in the stationary case, they represent a scale-wise decomposition of the process variance, i.e. $\var(S_t) = \sum_{j=1}^\infty \var(\wt{W}_{j,t})$ (see e.g. \cite{serroukh2000statistical}). As a result, the variance can also be well approximated by
\begin{equation}
    \var(S_t) \approx \sum_{j=1}^J \var(\wt{W}_{j,t}),
    \label{eq.var.decomp}
\end{equation}
since, for stationary processes and $J$ sufficiently large, the WV is close to zero for all $j > J$. Indeed, the WV is commonly finite for all levels for both stationary and non-stationary processes (or more precisely intrinsically stationary processes). Therefore, the WV provides a generalization of process variance through a scale-wise variance decomposition which also has many physical interpretations (as highlighted in the previous section and further on).

Considering this, we would therefore aim to minimize the variance of the virtual gyroscope using the representation in \eqref{eq.var.decomp}. Let us therefore re-express the WV of the virtual gyroscope as a function of the original processes using \eqref{eq:virtual:sensor} to obtain the following:
\begin{equation*}
\begin{aligned}
    \var(\wt{W}_{j,t}) &= \var\left(\sum_{l = 0}^{L_{j} - 1} h_{j, l} \sum_{i=1}^p c_i X_{i;t-l}\right) \\
    &= \var\left(\sum_{i=1}^p c_i W_{i,j,t}\right),
\end{aligned}
\end{equation*}
where $(W_{i,j,t})$ are the wavelet coefficients for level $j$ on the $i$-th signal $(X_{i;t})$. Now, letting $\W_{j,t}:= [W_{1,j,t}, \hdots, W_{p,j,t}]\trans$, we can re-write this relation in matrix form as
\begin{equation*}
    \var(\wt{W}_{j,t}) = \var(\c\trans \W_{j,t}) = \c\trans \var(\W_{j,t}) \c .
    \label{eq.virtual.wv}
\end{equation*}
The WV of the virtual gyroscope has the above form where $\var(\W_{j,t})$ is a $p \times p$ matrix with diagonal elements representing the WV of the individual signals at level $j$. Moreover, its off-diagonal elements are given by the Wavelet Cross-CoVariance (WCCV) (see \cite{whitcher2000wavelet, xu2019multivariate}) between the $i$-th and the $k$-th signals at time-lag zero for level $j$ which is defined as 
\begin{equation*}
    \gamma_{i,k,j} \vcentcolon = \text{Cov}(W_{i,j,t}, W_{k,j,t}).
\end{equation*}
One can easily notice that the WV is a special case of the WCCV when $i=k$. We can refer to $\A_j := \var(\W_{j,t})$ as the WCCV matrix at level $j$. Thus, as a result of the above developments and of the representation in \eqref{eq.var.decomp}, we can express the variance (or its generalization) of the virtual gyroscope as
\begin{equation}
    \sum_{j=1}^J \var(\wt{W}_{j,t}) = \sum_{j=1}^J \c\trans \A_j \c.
    \label{eq.quadratic.wv}
\end{equation}
However, in practice one may be interested in reducing the variance of the virtual gyroscope over certain time-scales (e.g. minimize short-run or long-run variance) and for this purpose we introduce the weights $\bomega = (\omega_1, \ldots, \omega_J)\trans \in \bOmega$ to each of the $J$ scales, where
\begin{equation*}
    \bOmega \vcentcolon = \left\{\bomega\in\real^J: \bomega\trans\1_J = 1 \;\; \text{and} \;\; \omega_j \geq 0 \;\; \text{for all} \;\; j \right\}.
\end{equation*}
For example, one may choose equal weights for all levels, implying that the minimization of the variance of $(S_t)$ occurs evenly across all time-scales. Nevertheless when it comes to navigation, as mentioned earlier one may be interested in maximizing navigation performance, i.e. minimizing the variance of $(S_t)$, over specific time-scales due to the nature of the navigation filter.

Taking into account the weights to the time-scales (i.e. levels), we can approximate $\var(S_t)$ with 
\begin{equation}
    \sum_{j=1}^J \c\trans \omega_j \A_j \c = \c\trans \underbrace{\sum_{j=1}^J \omega_j \A_j}_{\A_0} \c.
    \label{eq.final.form.var}
\end{equation}
The matrix $\A_0 \vcentcolon = (\sum_{j=1}^J \omega_j \gamma_{i,k,j})_{i,k=1,\ldots,p}$ therefore represents a weighted average of the WCCV matrices across all levels. Since all $\A_j$ are positive definite and all $\omega_j$ are non-negative, we have
\begin{equation*} 
    \det\left(\A_0\right) \geq \sum_{j = 1}^J \omega_j^p \det\left( \A_j\right) > 0,
\end{equation*}
implying that the matrix $\A_0$ is nonsingular. Based on the expression in \eqref{eq.final.form.var} we can finally define the optimal coefficient vector $\c_0$ as
\begin{equation}
\label{def:true_coef}
    \c_0 \vcentcolon = \underset{\c \in \boldsymbol{\mathcal{C}}}{\argmin}\; \c\trans \A_0 \c.
\end{equation}
Solving this minimization can be tricky due to the restriction of $\c\in\boldsymbol{\mathcal{C}}$. So alternatively, to respect these constraints, we can represent the above minimization problem through the use of a Lagrange multiplier $\lambda$ which modifies \eqref{def:true_coef} as follows
\begin{equation*}
    [\c_0\trans, \lambda_0]\trans \vcentcolon = \underset{\c\in\real^p, \lambda\in\real}{\argmin}\; \c\trans \A_0 \c + \lambda(\c\trans\1_p-1).
\end{equation*}
By setting the first derivative to zero, we can obtain the following well-known result: 
\begin{equation}
\label{eq:theo_closed_form}
    \c_0 = \frac{\A_0^{-1}\1_p}{\1_p\trans \A_0^{-1} \1_p}.
\end{equation}
With the closed-form expression, this solution can therefore be computationally efficient to obtain. However, we do not directly observe the matrix $\A_0$ so we replace it with a sample version $\wh{\A}$ defined as
$$\wh{\A}\vcentcolon=\left(\sum_{j=1}^J \omega_j \wh{\gamma}_{i,k,j}\right)_{i,k=1,\ldots,p},$$
which is an unbiased estimator of $\A_0$ where 
$$\wh{\gamma}_{i,k,j} \vcentcolon = \frac{1}{M_j} \sum_{t=1}^{M_j} W_{i,j,t} W_{k,j,t},$$ 
is the WCCV estimator for level $j$ and time-lag zero between the $i$-th and $k$-th signals, with $M_j \vcentcolon =T-L_j+1$ representing the number of wavelet coefficients generated at level $j$ with the maximum-overlap transform \cite{percival2006wavelet}. As a result of \cite{xu2019multivariate}, $\wh{\A}$ is a consistent estimator of $\A_0$, so $\wh{\A}$ is also nonsingular for sufficiently large sample sizes. It must also be underlined that the time indices for the above estimator should formally be $t = L_j, \hdots, T$, but for this work we assume a simple re-indexing of the wavelet coefficients to $t=1,\hdots, M_j$. Consequently, we deliver the following estimator for the optimal coefficients:
\begin{equation}
\label{eqn:c_hat_closed_form}
    \wh{\c} = \frac{\wh{\A}^{-1}\1_p}{\1_p\trans \wh{\A}^{-1} \1_p}.
\end{equation}
We call the approach based on the estimator $\wh{\c}$ the Scale-wise Variance Optimization (SVO).

\section{Statistical Properties}
\label{sec:main:res}

To study the statistical properties of the estimator $\wh{\c}$ we closely follow and adapt the discussion in \cite{xu2019multivariate} from which we borrow various definitions and conditions for the asymptotic properties of the WCCV estimator $\wh{\gamma}_{i,k,j}$. To start, we define
\begin{equation*}
    \bm{F}(\cdot) = \left(f^{(1)}(\cdot), \, f^{(2)}(\cdot), \, \dots, \, f^{(p)}(\cdot)\right)\trans,
\end{equation*}
as an $\real^p$-valued measurable function as well as the filtration $\mathcal{F}_t = \sigma(\dots, \, \epsilon_{t-1}, \, \epsilon_t)$ where $\epsilon_t$ are independently and identically distributed (i.i.d.) random variables. Then, we consider the first condition as follows:

\vspace{0.2cm}
\begin{enumerate}[label=\bfseries (C\arabic*), leftmargin=1cm]
    \item The multivariate process $(\W_{j,t})$ can be represented as \begin{equation*}
        \W_{j,t} = \bm{F}(\mathcal{F}_t).
    \end{equation*} 
    \label{cond:stationary}
\end{enumerate}
\vspace{-0.5cm}
This condition is commonly assumed to study the asymptotic properties of dependent processes and, if not directly stated, is often implied by other assumptions. For example, this condition is typically satisfied when the multivariate process $(\W_{j,t})$ is ergodic and (strictly) stationary implying that the processes $(W_{i,j,t})$, $(W^2_{i,j,t})$ and $(W_{i,j,t}W_{k,j,t})$ are stationary for all $i,k=1,\ldots,p$. This condition is respected for a wide class of time series models such as causal ARMA models and various state-space models (see e.g. \cite{wu2005nonlinear} and \cite{zhang2017gaussian}). 

We now consider an additional definition consisting in 
$$\W_{j,t}^{*} \vcentcolon = \bm{F}(\mathcal{F}_t^{*}),$$
with $\mathcal{F}_t^{*} = \sigma(\dots, \, \epsilon_0^{*}, \, \dots, \, \epsilon_{t-1}, \, \epsilon_t)$, where $\epsilon_0^{*}$ is an independent copy of $\epsilon_0$ from the same distribution. Following this definition, underlining that a consequence is that $\W_{j,t}^{*} = \W_{j,t}$ when $t < 0$, we can state the remaining conditions: \vspace{0.2cm}
\begin{enumerate}[label=\bfseries (C\arabic*), leftmargin=1cm, resume*]
    \item $\underset{i=1,\ldots,p}{\max} \mathbb{E}(W_{i,j,t}^4)^{1/4}  < \infty$. \label{cond:mom_bound} 
    \item $\underset{i=1,\ldots,p}{\max} \sum_{t=0}^{\infty} \mathbb{E}[(W_{i,j,t} - W_{i,j,t}^{*})^4]^{1/4} < \infty$. \label{cond:stability} 
\end{enumerate}
\vspace{0.2cm}
Both Conditions \ref{cond:mom_bound} and \ref{cond:stability} require the fourth moment of the wavelet coefficients (or a certain transformation of it) to be finite for all $i=1,\ldots,p$ and all $j = 1, \hdots, J$. In particular, Condition \ref{cond:stability} ensures that the cumulative impact of $\epsilon_0^{*}$ on the future values of the process $(W_{i,j,t})$ is finite, and therefore, it can be interpreted as a short-range dependence condition \citep{wu2011asymptotic}.

\begin{Remark}
As underlined in \cite{guerrier2021robust}, when using a Daubechies filter (such as the Haar wavelet filter) the wavelet coefficients $(W_{i,j,t})$ can be represented as a linear combination of the $d$-th order difference\footnote{Defining the backshift operator $B$ as the operator that admits the following equality $X_{i;t-1} = B X_{i;t}$, then the $d$-th order difference is given by $(1-B)^d X_{i;t}$.} of the original process $(X_{i;t})$, which we denote as $(\Delta_{i;t})$. In this case Conditions \ref{cond:stationary} to \ref{cond:stability} can be directly applied to $(\Delta_{i;t})$ instead of on $(W_{i,j,t})$.
\end{Remark}
\vspace{0.2cm}
In addition to the wide range of processes considered in \cite{wu2005nonlinear} and \cite{zhang2017gaussian}, using the representation considered in the above remark with the Haar wavelet filter, Appendix~\ref{proof:verify_cond_rw_ar1} verifies Conditions \ref{cond:stationary} to \ref{cond:stability} for a composite stochastic process made of a random walk and a finite sum of first-order autoregressive (AR(1)) processes, which is a commonly assumed process for inertial sensor error signals (see e.g. \cite{stebler2014generalized} and the references therein).

We now have all the required conditions to study the asymptotic properties of the estimator $\wh{\c}$ defined in \eqref{eqn:c_hat_closed_form}. Before addressing these properties, to ease notation we also define the following quantities:
$$\bg_0 \vcentcolon = (\gamma_{i,k,j})_{\substack{j=1,\ldots,J\\i,k=1,\ldots,p }},$$
and 
$$\wh{\bg} \vcentcolon = (\wh{\gamma}_{i,k,j})_{\substack{j=1,\ldots,J\\i,k=1,\ldots,p}},$$
which respectively represent the true and the estimated WCCV vectors. With these final definitions, the first property that we study is the consistency of the proposed estimator $\wh{\c}$ which is stated in the following theorem (followed by the proof).

\begin{Theorem}
\label{thm:consistency}
    Under Conditions \ref{cond:stationary} to \ref{cond:stability} we have that
    $$\|\wh{\c} - \c_0\|_2 = o_{\rm p}(1).$$
\end{Theorem}
\vspace{0.3cm}

\begin{proof}
From \cite{xu2019multivariate}, under Conditions \ref{cond:stationary} to \ref{cond:stability}, we have that $\wh{\bg} \overset{P}{\to} \bg_0$. By the continuous mapping theorem and the continuity of the matrix inversion, we have that $\wh{\A}^{-1}\1_p \overset{P}{\to} \A_0^{-1}\1_p$ and $\1_p\trans \wh{\A}^{-1}\1_p \overset{P}{\to} \1_p\trans\A_0^{-1}\1_p$. Therefore, we have that
\begin{equation*}
    \wh{\c} = \frac{\wh{\A}^{-1}\1_p}{\1_p\trans \wh{\A}^{-1}\1_p} \overset{P}{\to} \frac{\A_0^{-1}\1_p}{\1_p\trans \A_0^{-1}\1_p} = \c_0,
\end{equation*}
which concludes the proof. 
\end{proof}
\vspace{0.3cm}

As a consequence of Theorem \ref{thm:consistency}, whose proof is straightforward based on the results in \cite{xu2019multivariate}, we have confirmed that the proposed estimator indeed asymptotically targets the desired optimal coefficient vector $\c_0$. In order to perform inference (e.g. to understand if all signals have similar precision based on the value of $\wh{\c}$), we also need to derive the asymptotic distribution of $\wh{\c}$. To do so we further define $\V$ as the $Jp^2 \times Jp^2$ asymptotic covariance matrix of $\wh{\bg}$ as well as the $p \times Jp^2$ matrix $\G$ whose $i$-th column is given by 
\begin{equation*}
        \left[\frac{\I_p}{\1_p\trans\A_0^{-1}\1_p}  - \frac{\A_0^{-1}\1_p\1_p\trans}{(\1_p\trans\A_0^{-1}\1_p)^2} \right]\left(-\A_0^{-1} \frac{\partial \A_0}{\partial \gamma_i} \A_0^{-1} \1_p\right),
\end{equation*}
where $\gamma_i$ denotes the $i$-th element of a general WCCV vector $\bg$, and $\I_p$ is the $p \times p$ identity matrix. Using this notation, the following theorem states the asymptotic distribution of $\wh{\c}$ (followed by the proof).

\begin{Theorem}
\label{thm:asymp_norm}
    Under Conditions \ref{cond:stationary} to \ref{cond:stability}, we have that
    \begin{equation*}
        \sqrt{T}(\wh{\c} - \c_0) \overset{D}{\to} \mathcal{N}(\0, \bSigma),
    \end{equation*}
    where $\bSigma \vcentcolon = \G\V\G\trans$.
\end{Theorem}
\vspace{0.3cm}

\begin{proof}
For simplicity we define $\g(\bg): \real^{Jp^2} \to \boldsymbol{\mathcal{C}}$ as
\begin{equation*}
    \g(\bg)\vcentcolon= \frac{\A(\bg)^{-1} \1_p}{\1_p\trans \A(\bg)^{-1} \1_p},
\end{equation*}
where $\A(\bg) \vcentcolon = (\sum_{j=1}^J \omega_j \gamma^*_{i,k,j})_{i,k=1,\ldots,p}$ is the WCCV matrix of the general WCCV vector
$$\bg \vcentcolon = (\gamma^*_{i,k,j})_{\substack{j=1,\ldots,J\\i,k=1,\ldots,p}}.$$ 
As a consequence of this notation, we have that $\A_0 = \A(\bg_0)$ and $\wh{\A} = \A(\wh{\bg})$, implying that $\c_0 = \g(\bg_0)$ and $\wh{\c} = \g(\wh{\bg})$. Using the chain-rule we have for $i=1,\ldots, Jp^2$ that
\begin{equation*}
    \frac{\partial \g(\bg)}{\partial \gamma_i} = \left[\frac{\partial}{\partial \A(\bg)^{-1} \1_p} \frac{\A(\bg)^{-1} \1_p}{\1_p\trans \A(\bg)^{-1} \1_p}\right] \, \frac{\partial}{\gamma_i} \A(\bg)^{-1} \1_p,
\end{equation*}
where 
\begin{equation*}
\begin{aligned}
    &\frac{\partial}{\partial \A(\bg)^{-1} \1_p} \frac{\A(\bg)^{-1} \1_p}{\1_p\trans \A(\bg)^{-1} \1_p} \\ 
    &= \frac{\I_p}{\1_p\trans\A(\bg)^{-1}\1_p}  - \frac{\A(\bg)^{-1}\1_p\1_p\trans}{\{\1_p\trans\A(\bg)^{-1}\1_p\}^2},
\end{aligned}
\end{equation*}
and 
\begin{equation*}
    \frac{\partial}{\partial \gamma_i} \A(\bg)^{-1} \1_p = -\A(\bg)^{-1} \frac{\partial \A(\bg)}{\partial \gamma_i} \A(\bg)^{-1} \1_p.
\end{equation*}
Therefore, by the multivariate delta method and using the property that $\sqrt{T}(\wh{\bg} - \bg_0) \overset{D}{\to} \mathcal{N}(\0, \V)$ from \cite{xu2019multivariate}, under Conditions \ref{cond:stationary} to \ref{cond:stability} we have that
\begin{equation*}
\begin{aligned}
    \sqrt{T}(\wh{\c} - \c_0) = \sqrt{T}\left\{\g(\wh{\bg}) - \g(\bg_0)\right\} \overset{D}{\to} \mathcal{N}\left(\0, \G \V \G\trans\right),
\end{aligned}
\end{equation*}
which concludes the proof.
\end{proof}
\vspace{0.3cm}

With the results of this section we have verified that, under standard regularity conditions, the proposed estimator is statistically appropriate and converges in probability to the coefficients $\c_0$ defining the optimal virtual gyroscope we put forward in this work. More specifically, Theorem \ref{thm:asymp_norm} also provides the form of the asymptotic covariance matrix of $\wh{\c}$ which can be used to deliver confidence intervals for the proposed estimator $\wh{\c}$ thereby allowing to determine, for example, whether the collected signals are long enough, whether an equally weighted average is adequate to build a virtual gyroscope or to understand whether we should use certain signals to deliver the virtual signal. One can easily obtain an estimator of the asymptotic covariance matrix $\bm{\Sigma}$ by using parametric bootstrap if an underlying model is assumed for the error signals. However, in the context of this work we do not assume an underlying parametric model for the individual signals and we would therefore need to rely on non-parametric techniques, such as the batched-mean estimator \citep{zhang2017gaussian} or the progressive batched-mean method generalized from the idea in \cite{kim2013progressive}. In the next section we therefore consider the solution provided by one of these non-parametric methods, namely the Moving Block Bootstrap (MBB) \cite{kunsch1989jackknife, liu1992moving}, and show that this approach adequately quantifies the variability of the proposed estimator with asymptotically adequate coverage property.

\section{Covariance Computation}
\label{sec:comp}

Let us denote the estimator of the covariance matrix as $\wh{\bm{\Sigma}}$ which, given the ``sandwich'' form provided in Theorem \ref{thm:asymp_norm}, can be represented as
\begin{equation*}
    \wh{\bSigma} \vcentcolon = \wh{\G} \wh{\V} \wh{\G}\trans,
\end{equation*}
where $\wh{\V}$ and $\wh{\G}$ are respectively estimators of $\V$ and $\G$. While $\wh{\G}$ can be easily obtained as a plug-in estimator by using $\wh{\A}$, the estimation of $\V$ needs to rely on other computational techniques. For this purpose, we focus on the MBB procedure.

Firstly let us recall that the number of wavelet coefficients is denoted as $M_j$ for each level $j$ of wavelet decomposition. 
Since $M_j$ decreases as the level $j$ increases and the MBB procedure needs to be applied to vectors of the same length, we first need to obtain the same number of wavelet coefficients at each level while preserving the WCCV at time-lag zero between signals. More specifically, since we have $M_J \leq M_j$ for all $j$, we can remove $D_j \vcentcolon= M_j - M_J$ wavelet coefficients from level $j$, which results in $M_J$ wavelet coefficients at all levels for all individual signals. Thus, for the $i$-th individual process, we obtain a $J\times M_J$ ``trimmed'' wavelet coefficient matrix denoted as $\Y^{(i)}$ with the $j$-th row represented as $(W_{i,j,1}, \ldots, W_{i,j,M_J})$.

Next we can build a sequence of time-overlapping blocks from the matrices $\Y^{(i)}$ with $i=1,\ldots,p$. More specifically, considering the block size as $l\in\mathbb{N}_+$ and $\Y^{(i)}_t$ as the $t$-th column of $\Y^{(i)}$, we can create the following time-overlapping block matrices:
\begin{equation*}
\begin{aligned}
    & (\Y_1^{(i)}, \ldots, \Y_l^{(i)}), \\
    & (\Y_2^{(i)}, \ldots, \Y_{l+1}^{(i)}), \\
    & \ldots \\
    & (\Y_{M_J-l+1}^{(i)}, \ldots, \Y_{M_J}^{(i)}),
\end{aligned}
\end{equation*}
where each block matrix is of dimension $J\times l$ and the optimal block size is $l = \mathcal{O}(T^{1/3})$ (see e.g. \cite{buhlmann2002bootstraps}). If $M_J$ is a multiple of $l$, then we can sample with replacement $B \vcentcolon = M_J/l$ blocks independently with replacement:
\begin{equation*}
    \Y^{(i)*} \vcentcolon = (\Y_{U_1}^{(i)}, \ldots, \Y_{U_1+l-1}^{(i)}, \ldots, \Y_{U_B}^{(i)}, \ldots, \Y_{U_B+l-1}^{(i)}),
\end{equation*}
where the block-starting points $U_1,\ldots,U_B$ follow an i.i.d. discrete Uniform$(1,M_J-l+1)$ and are the same for all $i=1,\ldots,p$. If $M_J$ is not a multiple of $l$, then we resample $B \vcentcolon = \floor{M_J/l}+1$ blocks, where $\floor{\cdot}$ is the floor function, and use only the first columns of the $B$-th block such that the resulting $\Y^{(i)*}$ matrix is of dimension $J\times M_J$. Thus, the matrix $\Y^{(i)*}$ consists of the resampled wavelet coefficients, and we denote $W_{i,j,t}^*$ as the $(j,t)$-th entry of $\Y^{(i)*}$ with $i=1,\ldots,p$, $j=1,\ldots,J$ and $t=1,\ldots,M_J$. It must be noticed that in this MBB resampling procedure, the wavelet coefficients at time $t$ are resampled jointly for all signals at all levels. 

Having obtained a resampled version $W_{i,j,t}^*$ of $W_{i,j,t}$ for $t=1,\ldots,M_J$, we need to complete each level $j$ by completing $W_{i,j,t}^*$ with $t=M_J+1,\ldots,M_j$. More precisely, for each $i$-th individual process at level $j$, we consider time-overlapping vectors of length $D_j$:
\begin{equation*}
\begin{aligned}
    & (W_{i,j,1}, \ldots, W_{i,j,D_j}), \\
    & (W_{i,j,2}, \ldots, W_{i,j,D_j+1}), \\
    & \ldots \\
    & (W_{i,j,M_j-D_j+1}, \ldots, W_{i,j,M_j}). \\
\end{aligned}
\end{equation*}
Considering these vectors, we sample a time-index $t_j$ independently from the discrete Uniform$(1,M_j-D_j+1)$ for all $j=1,\ldots,J$. Then for each level $j$, we define 
\begin{equation*}
    (W_{i,j,M_J+1}^*, W_{i,j,M_j}^*) \vcentcolon= (W_{i,j,t_j}, \ldots, W_{i,j,t_j+D_j-1}),
\end{equation*}
for all $i=1,\ldots,p$. Having obtained all resampled wavelet coefficients, we can compute a bootstrapped estimator $\wh{\bg}^*$ for $\bg_0$ that is given by
\begin{equation*}
    \wh{\bg}^* \vcentcolon = (\wh{\gamma}_{i,k,j}^*)_{\substack{j=1,\ldots,J\\i,k=1,\ldots,p }},
\end{equation*}
with
\begin{equation*}
    \wh{\gamma}_{i,k,j}^* \vcentcolon = \frac{1}{M_j} \sum_{t=1}^{M_j} W_{i,j,t}^* W_{k,j,t}^*.
\end{equation*}
Repeating the above steps for $H\in\mathbb{N}_+$ times, we can consequently obtain a sequence of bootstrapped estimators $\{\wh{\bg}^*_1, \ldots, \wh{\bg}^*_H\}$, based on which we can construct an estimator $\wh{\V}^*$ for $\V$ that is defined as
\begin{equation*}
    \wh{\V}^* \vcentcolon = \frac{T}{H} \sum_{h=1}^H (\wh{\bg}^*_h - \wh{\bg})(\wh{\bg}^*_h - \wh{\bg})\trans.
\end{equation*}
In order for the bootstrap approximation to be adequate, we need $H$ to be sufficiently large. The complete procedure to compute $\wh{\V}^*$ is summarized in Algorithm~\ref{algo:V}.

Once the estimator $\wh{\V}^*$ is obtained, we can finally deliver an estimator $\wh{\bSigma}^*$ for the matrix of interest $\bSigma$ that is given by 
\begin{equation}
\label{def:est_Sigma}
    \wh{\bSigma}^* \vcentcolon= \wh{\G} \wh{\V}^* \wh{\G}\trans.
\end{equation}
As an additional note, the matrix $\wh{\bSigma}^*$ could eventually be computed directly within the MBB procedure. However, there is a general interest to obtain an appropriate estimator for the matrix $\V$ in various applications as highlighted at the end of this section. Moreover, the use of a plug-in estimator $\wh{\G}$ possibly allows to diminish the dependence of the final estimator $\wh{\bSigma}^*$ on the resampling procedure. 

\begin{Remark}
In the case where $(\X_t)$ is jointly stationary, one can directly perform the MBB on $(\X_t)$ to compute $\wh{\bSigma}^*$ which would avoid the resampling step to ensure $W_{i,j,t}^*$ with $t=1,\ldots,M_j$ for all $j = 1, \hdots, J$.
\end{Remark}
\vspace{0.2cm}

\begin{algorithm}
\caption{Algorithm to compute $\wh{\V}^{*}$}
\label{algo:V}
\begin{algorithmic}[1]
\renewcommand{\algorithmicrequire}{\textbf{Input:}}
\renewcommand{\algorithmicensure}{\textbf{Output:}}
\REQUIRE $(W_{i,j,t}:i=1,\ldots,p; j=1,\ldots,J; t=1,\ldots,M_j)$
\ENSURE  $\wh{\V}^{*}$
\\ \STATE Trim the wavelet coefficients $(W_{i,j,t})$ to obtain matrices $\Y^{(i)}$ of dimension $J\times M_J$ for all $i=1,\ldots,p$.
\\ \STATE Create blocks $(\Y_k^{(i)}, \ldots, \Y_{k+l-1}^{(i)})$ with $k=1,\ldots,M_j-l+1$ for all $i$.
\\ \STATE Create vectors $(W_{i,j,k},\ldots,W_{i,j,k+D_j-1})$ with $k=1,\ldots,M_J-D_j+1$ for all $i$ and $j$.
\\ 
\FOR {$h = 1$ to $H$}
\STATE Sample independently with replacement block-starting points $U_1,\ldots,U_B$ from discrete Uniform$(1,M_J-l+1)$, based on which we construct $\Y^{(i)*}$ to obtain $W_{i,j,t}^*$ for all $i,j$ with $t=1,\ldots,M_J$.
\STATE Sample vector-starting points $t_j$ independently from discrete Uniform$(1,M_j-D_j+1)$ for all $j$ and concatenate the sampled vector to the end of $W_{i,j,t}^*$ for all $i,j$.
\STATE Compute $\wh{\bg}_h^*$ using the bootstrapped wavelet coefficients $(W_{i,j,t}^*: i=1,\ldots,p; j=1,\ldots,J; t=1,\ldots,M_j)$.
\ENDFOR
\STATE Compute $\wh{\V}^*$.\RETURN $\wh{\V}^*$ 
\end{algorithmic} 
\end{algorithm} 

Given the proposed procedure, we want to verify that the considered estimator $\wh{\bSigma}^*$ delivers adequate (asymptotic) inference allowing to quantify sampling uncertainty in the estimation of the optimal coefficients $\c_0$. More specifically, setting a significance level $\alpha \in (0,1)$, we would want the following interval to asymptotically include $\c_0$ with probability $1 - \alpha$:
\begin{equation}
\label{def:coef_CI}
    \left(\wh{c}_i \pm z_{1-\alpha/2} \sqrt{\wh{\Sigma}^*_{i,i}/T}\right),
\end{equation}
where $z_{1-\alpha/2}$ represents the $(1-\alpha/2)$-quantile of the standard normal distribution and $\wh{\Sigma}^*_{i,i}$ is the $(i,i)$-th entry of the matrix $\wh{\bSigma}^*$. The reason for considering this interval lies in the asymptotic behavior of the estimator $\wh{\c}$ which, as stated in Theorem~\ref{thm:asymp_norm}, is asymptotically normally distributed with covariance matrix $\bSigma$. Therefore, denoting $c_{0,i}$ as the $i$-th element of $\c_0$ and $\gamma_{0,i}$ as the $i$-th element of $\bg_0$, Theorem~\ref{thm:block_boot} below states the validity of the estimator $\wh{\bSigma}^*$ in achieving asymptotically accurate coverage. A summary proof is provided immediately after the stated theorem while a more detailed proof can be found in Appendix~\ref{proof:block_boot}.

\begin{Theorem}
\label{thm:block_boot}
Under Conditions \ref{cond:stationary} to \ref{cond:stability}, for $i=1,\ldots,p$, we have
\begin{equation*}
    \lim_{T\to\infty} \Pr\left[c_{0,i} \in \left(\wh{c}_i \pm z_{1-\alpha/2} \sqrt{\wh{\Sigma}^*_{i,i}/T}\right)\right] = 1-\alpha.
\end{equation*}
\end{Theorem}
\vspace{0.2cm}

\begin{proof}
To show Theorem~\ref{thm:block_boot}, we essentially need to show that 
\begin{equation*}
    \sqrt{T}\wh{\bm{\Sigma}}^{*\, -1/2} (\wh{\c} - \c_0) \overset{D}{\to} \mathcal{N}(\0, \mathbf{I}_p),
\end{equation*}
where the left hand side can also be expressed as 
\begin{equation*}
    \sqrt{T}\bm{\Sigma}^{-1/2} (\wh{\c} - \c_0) + \sqrt{T}\left(\wh{\bm{\Sigma}}^{*\, -1/2} -  \bm{\Sigma}^{-1/2} \right) (\wh{\c} - \c_0).
\end{equation*}
As a result of Theorem~\ref{thm:asymp_norm} we have that 
\begin{equation*}
    \lim_{T \to \infty}\,  \Pr\left[c_{0,i} \in \left(\wh{c}_i \pm z_{1-\alpha/2} \sqrt{\Sigma_{i,i}/T}\right)\right] = 1-\alpha,
\end{equation*}
thus providing the desired properties for the first term in the previous expansion. Therefore, we are left with determining the asymptotic behavior of the second term, i.e. $\sqrt{T}(\wh{\bm{\Sigma}}^{*\, -1/2} -  \bm{\Sigma}^{-1/2}) (\wh{\c} - \c_0)$. Since $\wh{\bm{\Sigma}}^{*} = \wh{\G} \wh{\V}^* \wh{\G}\trans$, we want to determine the asymptotic behavior of $\wh{\V}^*$ and $\wh{\G}$. Firstly, based on Conditions \ref{cond:stationary} to \ref{cond:stability}, by the results in \cite{buhlmann2002bootstraps} and using Cauchy-Schwarz and Markov's inequalities, through some computations we can obtain
\begin{equation*}
    \wh{V}^*_{i,k} = V_{i,k} + \mathcal{O}_{\rm p}(T^{-1/3}),
\end{equation*}
while by applying the multivariate delta method we can obtain $\wh{G}_{i,k} = G_{i,k} + \mathcal{O}_{\rm p}(T^{-1/2})$. By combining these results we can obtain that
\begin{equation*}
    \sqrt{T}\left(\wh{\bm{\Sigma}}^{*\, -1/2} -  \bm{\Sigma}^{-1/2} \right) (\wh{\c} - \c_0) = o_{\rm p}(1),
\end{equation*}
which allows us to obtain
\begin{equation*}
    \lim_{T \to \infty}\,  \Pr\left[c_{0,i} \in \left(\wh{c}_i \pm z_{1-\alpha/2} \sqrt{\wh{\Sigma}^*_{i,i}/T}\right)\right] = 1-\alpha,
\end{equation*}
thus concluding the proof.
\end{proof}
\vspace{0.2cm}

\begin{Remark}
The result of Theorem \ref{thm:block_boot} can be refined if the distribution of $\sqrt{T}\wh{\Sigma}_{i,i}^{*\,-1/2}(\wh{c}_i - c_{0,i})$ admits an Edgeworth expansion, a requirement that is generally satisfied under suitable regularity conditions (usually moment and smoothness conditions, see \cite{hall2013bootstrap} and the references therein). Indeed, using Theorem \ref{thm:block_boot}, we have 
\begin{equation*}
    \sqrt{T}\wh{\Sigma}_{i,i}^{*\,-1/2}(\wh{c}_i - c_{0,i}) \overset{D}{\to} \mathcal{N}(0, 1).
\end{equation*}
Applying the Edgeworth expansion on the cumulative distribution function of $\sqrt{T}\wh{\Sigma}_{i,i}^{*\,-1/2}(\wh{c}_i - c_{0,i})$ and under additional regularity requirements, we can obtain
\begin{equation*}
\sup_{z \in \real} \; \left|\Pr\left[\sqrt{T}\wh{\Sigma}_{i,i}^{*\,-1/2}(\wh{c}_i - c_{0,i}) < z\right] - \Phi(z) \right| < \frac{C}{\sqrt{T}},
\end{equation*}
where $C$ is some finite positive constant and $\Phi(\cdot)$ denotes the cumulative distribution function of the standard normal distribution. Therefore, the result of Theorem \ref{thm:block_boot} could be refined as follows:
\begin{equation*}
    \Pr\left[c_{0,i} \in \left(\wh{c}_i \pm z_{1-\alpha/2} \sqrt{\wh{\Sigma}^*_{i,i}/T}\right)\right] = 1-\alpha + \mathcal{O}\left(T^{-1/2}\right).
\end{equation*}
\end{Remark}
\vspace{0.2cm}

Theorem \ref{thm:block_boot} is therefore important for this work but, in addition, the results required to prove this theorem are essential for many other applications that make use of the WV. In order to highlight these results we state the following corollaries (and the corresponding proofs).

\begin{Corollary}
Under Conditions \ref{cond:stationary} to \ref{cond:stability}, we have 
\begin{equation*}
    \|\wh{\V}^* - \V\|_F = \mathcal{O}_{\rm p}(T^{-1/3}),
\end{equation*}
where $\|\cdot\|_F$ denotes the Frobenius norm.
\end{Corollary}
\vspace{0.2cm}

\begin{proof}
The proof is directly obtained using the result that for $i,k=1,\ldots,Jp^2$ we have $\wh{V}_{i,k}^* = V_{i,k} + \mathcal{O}_{\rm p}(T^{-1/3})$, as shown in the proof of Theorem~\ref{thm:block_boot}.
\end{proof}
\vspace{0.2cm}

\begin{Corollary}
Under Conditions \ref{cond:stationary} to \ref{cond:stability}, for $i=1,\ldots,Jp^2$ we have
\begin{equation*}
    \lim_{T\to\infty} \Pr\left[\gamma_{0,i} \in \left(\wh{\gamma}_i \pm z_{1-\alpha/2} \sqrt{\wh{V}^*_{i,i}/T}\right)\right] = 1-\alpha.
\end{equation*}
\end{Corollary}
\vspace{0.2cm}

\begin{proof}
Similarly to the proof of Theorem~\ref{thm:block_boot}, we want to show that 
\begin{equation*}
    \sqrt{T} \wh{\V}^{*\,-1/2} (\wh{\bg}-\bg_0) \overset{D}{\to} \mathcal{N}(\0, \I_{Jp^2}).
\end{equation*}
From the proof of Theorem~\ref{thm:block_boot} we have that $\wh{V}_{i,k}^*=V_{i,k}+\mathcal{O}_{\rm p}(T^{-1/3})$ for $i,k=1,\ldots,Jp^2$ and by the delta method we obtain $\wh{V}_{i,k}^{*\,-1/2}=V_{i,k}^{-1/2}+\mathcal{O}_{\rm p}(T^{-1/3})$. Therefore, since $\sqrt{T}(\wh{\bg}-\bg_0) \overset{D}{\to}\mathcal{N}(\0,\V)$ based on \cite{xu2019multivariate}, we have that
\begin{equation*}
\begin{aligned}
    & \sqrt{T} \wh{\V}^{*\,-1/2} (\wh{\bg}-\bg_0) \\ 
    &= \sqrt{T} \V^{-1/2} (\wh{\bg}-\bg_0) + \sqrt{T} (\wh{\V}^{*\,-1/2} - \V^{-1/2}) (\wh{\bg}-\bg_0) \\
    &= \sqrt{T} \V^{-1/2} (\wh{\bg}-\bg_0) + \mathcal{O}_{\rm p}(T^{-1/3}) \overset{D}{\to} \mathcal{N}(\0, \I_{Jp^2}),
\end{aligned}
\end{equation*}
thus concluding the proof.
\end{proof}
\vspace{0.2cm}

As mentioned earlier, a non-parametric estimator of the asymptotic covariance matrix $\V$ is of great importance for all applications that perform inference based on this quantity. Among others, examples of such applications are related to Portmanteau tests, model estimation and selection (see e.g. \cite{guerrier2013wavelet, guerrier2021robust, gallegati2012wavelet, xie2013wavelet, jia2015correlations, gencay2015multi, Sanderson469} to mention a few). This approach is consequently used for the simulation and case studies presented in the following sections also highlighting its relevance in the context of this work.
 
\section{Simulation Studies}
\label{sec:study}
 
We present two simulation studies to give support to the theoretical properties of the SVO method and compare it with an alternative method put forward in~\cite{vaccaro2017reduced}, which, to the best of our knowledge, is currently the only available alternative method to {\color{black} optimize the stochastic properties of the resulting} virtual signal. The latter approach is based on a parametric assumption on the underlying process that generates the individual gyroscope error signals. Therefore, to make a fair comparison, in the first simulation (Section~\ref{sec:study_1}) we assume that the true stochastic model for the individual gyroscopes is the same as the one considered in \cite{vaccaro2017reduced}. Considering the complexity of stochastic signals issued from gyroscopes, this model is relatively simple but often accepted for high-grade sensors. In the second simulation (Section~\ref{sec:study_2}), the stochastic model considered is substantially more complex and more representative of low-cost MEMS sensors. In that case it would be considerably more difficult to formulate an optimal signal combination based on a parametric assumption. In this setting, which essentially better approximates the complexity of stochastic error signals in real application scenarios, it is possible to observe the potential advantages of the non-parametric method presented in this paper. 

Since the goal of these simulation studies is to understand how the SVO method delivers virtual signals that reduce the WV of future (out-of-sample) signal arrays, both studies are performed as follows:
\begin{enumerate}
    \item Simulate 50 signal arrays (each with six signals) from a fixed model.
    \item For each array, we apply each method to obtain optimal coefficients, delivering 50 vectors of optimal coefficients for each method.
    \item For each coefficient vector, we simulate 10 new signal arrays (each with six signals) from the same model and compute the optimal virtual signal for each signal array based on the respective coefficient vector, thus resulting in 10 virtual signals for each coefficient vector.
\end{enumerate}
Overall, using independent array samples from the same model for each step, the above procedure generates 500 out-of-sample virtual signals for each method. In all cases, the individual signals that compose the arrays have a length that is equivalent to 29 hours recording at 10 Hz, allowing us to have a maximum of 19 levels of wavelet decomposition which we take fully advantage of by choosing $J = 19$.

\subsection{Case 1}
\label{sec:study_1}

The setting used in \cite{vaccaro2017reduced} consisted in an array of six gyroscopes where all stochastic error signals follow a fixed composite model composed by a white noise and a random walk defined as follows: 
\begin{equation}
\begin{aligned}
\label{eq:gyromodel1}
    \mathbf{X}_t & = \1_6 \delta + \mathbf{b}_t + \bm{\xi}_t, & \bm{\xi}_t \simiid \mathcal{N}(\mathbf{0}, \mathbf{R}), \\
    \mathbf{b}_t & = \mathbf{b}_{t-1} + \bm{\eta}_t, & \bm{\eta}_t \simiid \mathcal{N}(\mathbf{0}, \mathbf{Q}),
\end{aligned}
\end{equation}
where $\delta$ is the constant angular velocity described in Section~\ref{sec:syn:sensors}. Here the matrix $\mathbf{R}$ is diagonal, implying that the components of the white noise are independent for a given time $t$, where the diagonal elements are given by
\begin{equation}
\label{eqn:diag_R}
    [2.806, 1.978, 1.361, 1.064, 1.069, 2.539],
\end{equation}
%
with the values expressed in $\text{deg}^2/\text{s}^2 \times 10^{7}$. On the other hand, as often observed in practice, the innovations of the random walk are correlated at a given time $t$, with diagonal elements of the covariance matrix $\mathbf{Q}$ given by
\begin{equation}
\label{eqn:diag_Q}
    [0.255, 0.472, 3.489, 2.092, 0.555, 2.501],
\end{equation}
with the values expressed in $\text{deg}^2/\text{s}^2 \times 10^{13}$. For ease of illustration, the intensity of the correlation matrix $\mathbf{Q}$ is graphically represented in Figure~\ref{fig:s001_corr} while its exact values are given in Appendix~\ref{app:simparams:case1}. We highlight that in \cite{vaccaro2017reduced} (Section 2.9) the above reported quantities are expressed as a function of the continuous noise power spectral density whereas in this work they are presented in discrete time units.
 
\begin{figure}[h!]
    \centering
    \includegraphics{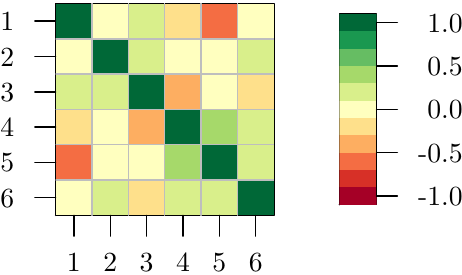}
    \caption{Absolute correlation matrix $\mathbf{Q}$ of the random walk innovations for the composite process considered in the simulation presented in Section \ref{sec:study_1}.}
    \label{fig:s001_corr}
\end{figure}

It is not possible to make a direct comparison between the SVO method and the approach put forward in \cite{vaccaro2017reduced}. Indeed, the latter is fully parametric and explicitly aims at minimizing the variance of the random walk innovations of the resulting virtual gyroscope, whereas the SVO method, as stated earlier, is non-parametric and consequently makes no parametric assumptions on the underlying signal-generating process. More specifically, as derived from the definitions in \cite{vaccaro2017reduced} with $\delta = 0$, the virtual gyroscope signal defined as a linear combination of individual gyroscopes that satisfy \eqref{eq:gyromodel1} can be expressed as follows:
\begin{equation*}
\begin{aligned}
    & V_t \vcentcolon = \c\trans \X_t = \wt{b_t} + \wt{\xi_t}, \\
    & \wt{b_t} \vcentcolon = \c\trans \b_t = \wt{b}_{t-1} + \wt{\eta}_t,
\end{aligned}
\end{equation*}
with
\begin{equation*}
\begin{aligned}
    & \wt{\xi_t} \vcentcolon = \c\trans \bm{\xi_t} \simiid \mathcal{N}(0, \sigma^2), \quad \sigma^2 \vcentcolon = \c\trans \mathbf{R}\c, \\
    & \wt{\eta}_t \vcentcolon = \c\trans\bm{\eta}_t \simiid \mathcal{N}(0, \gamma^2), \quad \gamma^2 \vcentcolon = \c\trans\mathbf{Q}\c.
\end{aligned}
\end{equation*}
For the SVO method we present its performance based on two different weight vectors on time-scales denoted respectively as $\bomega_1$ and $\bomega_2$, the actual values of which are provided in Appendix~\ref{app:simparams:case1}. The first weight vector $\bomega_1$ aims at assigning more weights to the higher levels of decomposition, where the impacts of the variance of the random walk process are noticeable. Hence, this weight vector is closely related to the coefficient vector proposed by \cite{vaccaro2017reduced} which we denote as $\wh{\mathbf{c}}_\text{RDVG}$. On the other hand, the second weight vector $\bomega_2$ attempts to achieve the opposite goal, namely to reduce the variability of the virtual gyroscope at the lowest scales with the aim to reduce the variance of the white noise process. The latter weight vector is therefore chosen to highlight the flexibility of the proposed SVO method. The coefficients based on $\bomega_1$ and $\bomega_2$ using the SVO method are denoted respectively as $\wh{\mathbf{c}}_{\bomega_1}$ and $\wh{\mathbf{c}}_{\bomega_2}$.

\begin{figure}[t]
    \centering
    \includegraphics{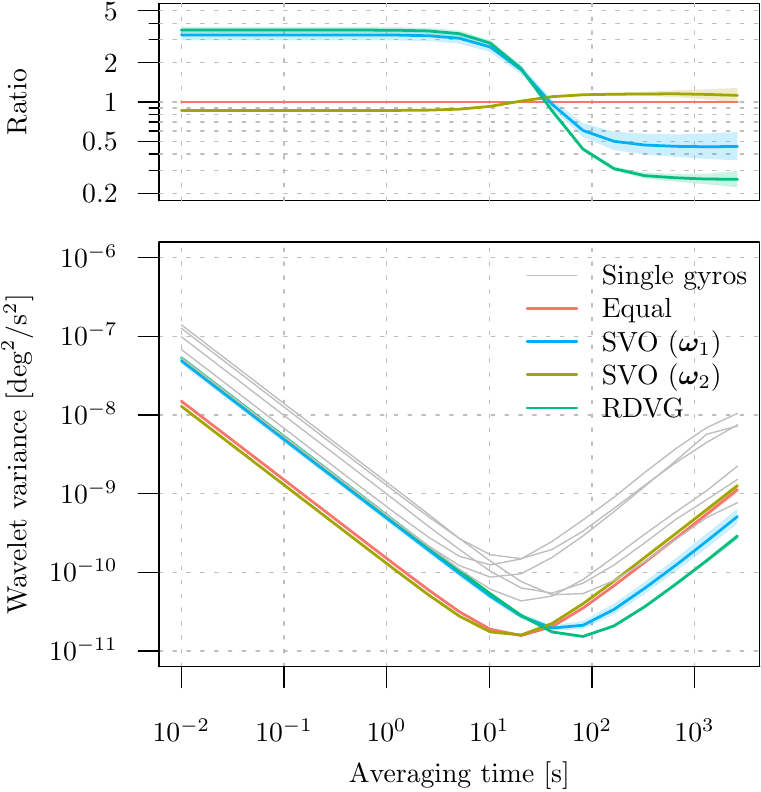}
    \caption{Lower plot: Comparison of the WV of the virtual gyroscopes obtained with different methods considered in the simulation presented in Section~\ref{sec:study_1}. Upper plot: Summary of the ratios of the WV of the virtual gyroscopes based on each method with the WV of the virtual gyroscope computed based on the equally weighted coefficients $\mathbf{c}_{\text{Eq}}$.}
    \label{fig:case1wv}
\end{figure}

We first compare the performance of these methods in terms of the WV of the constructed virtual signals in Figure~\ref{fig:case1wv}. More precisely, we compare the WV of the virtual signals computed using the SVO method based on both $\bomega_1$ and $\bomega_2$, as well as the one obtained with the approach of \cite{vaccaro2017reduced}. As a reference, we also present the WV of the virtual signals delivered by equally weighted coefficients (i.e. $\mathbf{c}_{\text{Eq}} \vcentcolon= 1/6 \cdot\1_6$). In the lower plot of Figure~\ref{fig:case1wv} we present the average WV of the 500 out-of-sample virtual signals obtained with each method in this simulation. We also present the confidence intervals of these averages for each method considered, as well as the WV of one randomly selected out-of-sample array of signals from the simulation to visualize how all methods allow to reduce the WV compared to the original array signals. As we can see from this lower plot, the WV of the virtual signals based on $\mathbf{c}_{\text{Eq}}$ and $\wh{\mathbf{c}}_{\bomega_2}$ appear extremely close, and nevertheless the latter has the lowest WV at the first scales among all methods. This is expected as the weight vector $\bomega_2$ is chosen to minimize the WV of the virtual signals at the short time-scales. On the other hand, the method based on \cite{vaccaro2017reduced} ($\wh{\mathbf{c}}_\text{RDVG}$) and the SVO method based on the first set of weights ($\wh{\mathbf{c}}_{\bomega_1}$) appear to minimize the WV of the virtual signals at the large time-scales as expected by construction. However, the approach based on \cite{vaccaro2017reduced} appears to decrease the WV of the virtual signals more than the SVO method based on $\bomega_1$ at the large time-scales. This is also expected as the former is exactly based on the considered parametric model that generates the individual signals and aims at reducing the variance of the random walk process which dominates these larger time-scales. The upper plot of Figure~\ref{fig:case1wv} further summarizes the comparison of the different approaches by representing the ratios of the WV of the virtual signals based on each method with the WV of the virtual signals based on $\mathbf{c}_{\text{Eq}}$. We can again observe from this plot that the WV of the virtual signals based on $\wh{\mathbf{c}}_\text{RDVG}$ and $\wh{\mathbf{c}}_{\bomega_1}$ are much larger at the short time-scales, which can be reduced using $\wh{\mathbf{c}}_{\bomega_2}$.

\begin{figure}[t]
    \centering
    \includegraphics{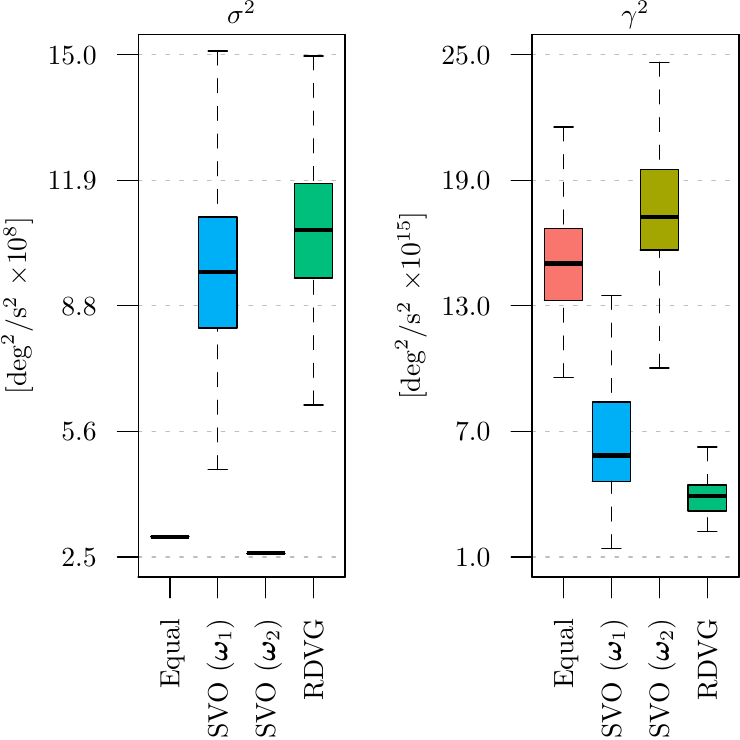}
    \caption{Estimated values for the white noise parameter $\sigma^2$ and the random walk parameter $\gamma^2$ of the virtual gyroscopes based on different methods considered in the simulation presented in Section~\ref{sec:study_1}.}
    \label{fig:case1params}
\end{figure}

For this simulation we also compare the variances of the white noise and random walk processes by estimating $\sigma^2$ and $\gamma^2$ using the virtual signals obtained by each method. Indeed, the objective of the method in \cite{vaccaro2017reduced} is to minimize the variance of the random walk process (i.e., $\gamma^2$), so we expect the estimated $\gamma^2$ on the corresponding virtual signals to have a comparatively lower value. The empirical distributions of the estimated $\sigma^2$ and $\gamma^2$ for the virtual signals of each method are summarized in Figure~\ref{fig:case1params}, where the parameters are estimated using the Generalized Method of Wavelet Moments (GMWM) put forward in \cite{guerrier2013wavelet}. As can be seen, when analyzing the empirical distribution for the random walk parameter $\gamma^2$, the method proposed in \cite{vaccaro2017reduced} indeed delivers consistently lower estimates thereby confirming its optimality under this parametric setting. The equally weighted approach and the SVO method based on both $\bomega_1$ and $\bomega_2$ deliver higher estimated values of $\gamma^2$. Nevertheless, the SVO method based on $\bomega_1$ can achieve low estimated values of $\gamma^2$, comparable to the ones obtained by \cite{vaccaro2017reduced} but with greater variability. On the other hand, it can be seen that the equally weighted approach and the SVO method based on $\bomega_2$ greatly reduces the value of the white noise parameter $\sigma^2$ compared to the other two methods. Overall, one may conclude that the SVO method appears to have a reasonable performance compared to the approach put forward in \cite{vaccaro2017reduced}. 

\begin{figure}[t]
    \centering
    \includegraphics{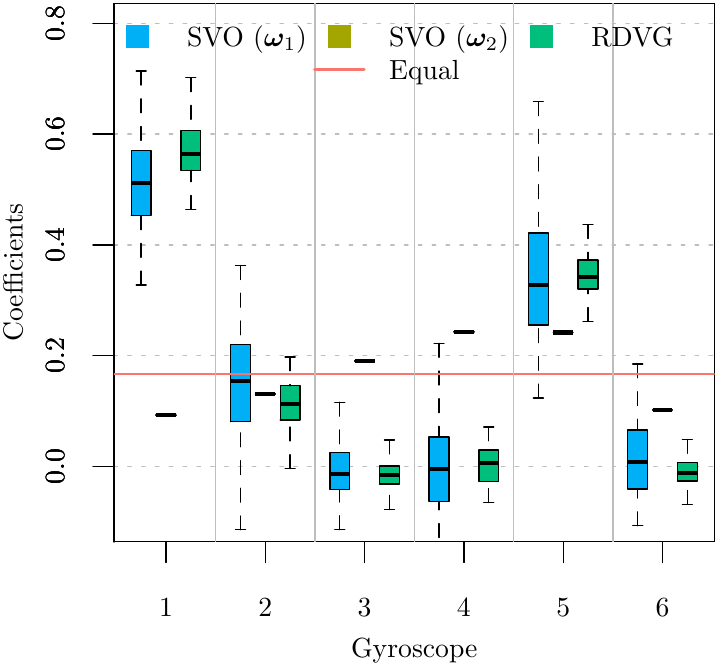}
    \caption{Empirical distributions of the estimated coefficients $\wh{\mathbf{c}}_\text{RDVG}$, $\wh{\mathbf{c}}_{\bomega_1}$, and $\wh{\mathbf{c}}_{\bomega_2}$ computed in the simulation presented in Section~\ref{sec:study_1}.}
    \label{fig:case1coeffs}
\end{figure}

An additional aspect that is interesting to consider is the empirical distributions of the estimated coefficients for each method, i.e. $\wh{\mathbf{c}}_\text{RDVG}$, $\wh{\mathbf{c}}_{\bomega_1}$ and $\wh{\mathbf{c}}_{\bomega_2}$, which are summarized in Figure~\ref{fig:case1coeffs}. We recall that 50 coefficient vectors are issued from this simulation for each method, implying that each boxplot represents 50 data-points. Although the diagonal elements of the model innovation covariance matrices $\mathbf{R}$ and $\mathbf{Q}$ (given in \eqref{eqn:diag_R} and \eqref{eqn:diag_Q} respectively) cannot be interpreted independently from the off-diagonal elements, it is interesting to see how the estimated coefficients appear to be consistent with the values of the innovation variances for each method. More specifically, we highlight that the first, second and sixth signals have the largest variances for the white noise process (as given in \eqref{eqn:diag_R}, with values of $2.806$, $1.978$ and $2.539$ $\text{deg}^2/\text{s}^2 \times 10^{7}$). The SVO method based on $\bomega_2$ which aims exactly at minimizing the virtual signal variance at the short time-scales obtains coefficients that are consistently lower than the equal weights $\c_\text{Eq}$ for these three individual signals. On the other hand, the third, fourth and sixth signals have the largest variances for the random walk process (as given in \eqref{eqn:diag_Q} with values of $3.489$, $2.092$ and $2.501$ $\text{deg}^2/\text{s}^2 \times 10^{13}$), and it can be seen that the computed $\wh{\mathbf{c}}_\text{RDVG}$ and $\wh{\mathbf{c}}_{\bomega_1}$ are smaller than $\c_\text{Eq}$ for these three individual signals. Therefore, Figure~\ref{fig:case1params} confirms that these methods are delivering intuitive coefficients under this perspective and highlights again the flexibility of the SVO method when employing different weights on time-scales. 

As a final result for this simulation, we consider how well the covariance computation method studied in Section~\ref{sec:comp} is able to deliver adequate confidence intervals for the coefficients. Since the last scales of WV are more variable
, we focus on studying the coverage of the confidence intervals based on the coefficients obtained using the SVO method with $\bomega_2$. Indeed, $\c_{\bomega_2}$ is less dependent on the last scales by construction and more on the first scales, which allows to better illustrate the asymptotic properties of the estimated coefficients, whose asymptotic covariance can be approximated by the proposed estimator $\wh{\bSigma}^*$ given in \eqref{def:est_Sigma}. Since the considered model has a closed-form representation of the WCCV, we can directly compute the matrix $\A_0$ defined in \eqref{eq.final.form.var} based on $\bomega_2$ and therefore obtain the true coefficients $\c_0$ with \eqref{eq:theo_closed_form}. We then run $10^3$ Monte-Carlo replications using the same model where, for each replication, we compute the 95\% confidence intervals given by \eqref{def:coef_CI} based on $\wh{\c}_{\bomega_2}$ and $\wh{\bSigma}^*$, and verify whether each element of $\c_0$ is captured in the respective confidence intervals.   

\begin{figure}[h]
    \centering
    \includegraphics{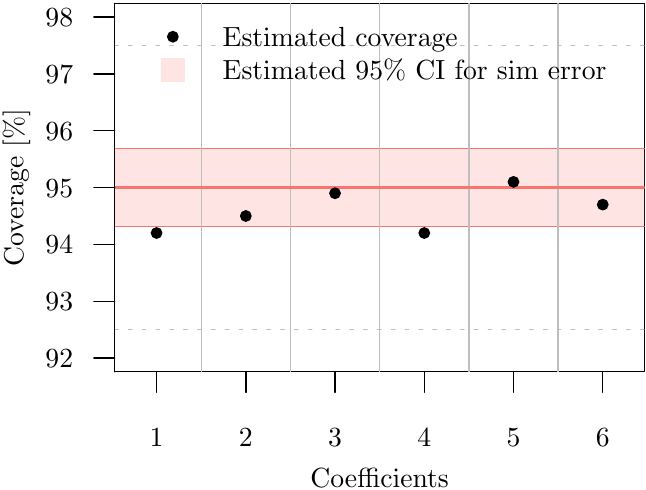}
    \caption{Empirical coverage of the 95\% confidence intervals computed based on $\wh{\c}_{\bomega_2}$ and $\wh{\bSigma}^*$ with $10^3$ Monte-Carlo replications for the simulation presented in Section~\ref{sec:study_1}.}
    \label{fig:coverage}
\end{figure}

Figure~\ref{fig:coverage} presents the coverage results based on the above procedure. In particular, the horizontal shaded area represents the estimation error for the empirical coverage, i.e. the region within which an estimated coverage would be considered accurate given the number of Monte-Carlo replications. As the dots represent the estimated coverage for each of the six coefficients in $\wh{\mathbf{c}}_{\bomega_2}$, we can see that all the empirical coverages lie within (or are extremely close to) the acceptable region and are therefore adequate. This additional result provides support to the validity of the asymptotic behaviour of the estimated coefficients and also of the proposed method to estimate $\bSigma$ described in Section \ref{sec:comp}.

To conclude, this simulation study highlights that the following points: (i) the approach put forward in \cite{vaccaro2017reduced} is the best among all considered methods in terms of reducing the large-scale variance of the chosen model which is not surprising knowing that it is tailored to this parametric setting while the SVO method is non-parametric and consequently does not benefit from any parametric information (nevertheless, overall the SVO method delivers good performance in terms of the WV of the constructed virtual signals as well as the estimated parameter values for $\sigma^2$ and $\gamma^2$); (ii) both the method of \cite{vaccaro2017reduced} and the SVO provide intuitive coefficients given the values of the innovation variances of individual signals; (iii) the SVO is more flexible than the approach of \cite{vaccaro2017reduced} as it can employ different weights on time-scales that can be tailored to different applications; (iv) the SVO can be used to construct confidence intervals on the coefficients and deliver accurate empirical coverage.

\subsection{Case 2}
\label{sec:study_2}

\begin{figure}[h!]
    \captionsetup[subfigure]{labelformat=empty}
     \centering
     \begin{subfigure}[b]{0.31\columnwidth}
         \centering
         \includegraphics[scale=1]{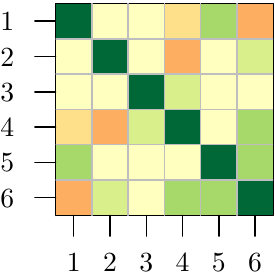}
         \caption{$\mathbf{P}_1$}
     \end{subfigure}
     \hfill
     \begin{subfigure}[b]{0.31\columnwidth}
         \centering
         \includegraphics[scale=1]{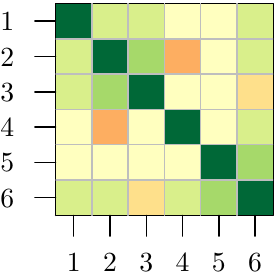}
         \caption{$\mathbf{P}_2$}
         \end{subfigure}
     \hfill
     \begin{subfigure}[b]{0.31\columnwidth}
         \centering
         \includegraphics[scale=1]{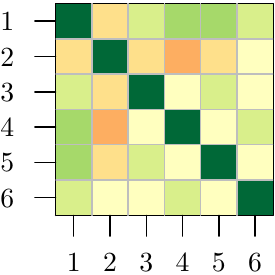}
         \caption{$\mathbf{P}_3$}
     \end{subfigure}
        \caption{Absolute correlation matrices of the AR(1) processes considered in the simulation presented in Section~\ref{sec:study_2}. (For the color scale please refer to Figure~\ref{fig:s001_corr}.)}
        \label{fig:s002_corr}
\end{figure}

In this section, we consider a case in which the underlying stochastic model of individual gyroscope errors is more complex, composed by the sum of a white noise process and three first-order Auto-Regressive, AR(1), processes, which are essentially reparametrizations of first-order Gauss-Markov processes. More precisely, setting the constant angular velocity $\delta = 0$, we define this composite process as
\begin{equation}
\begin{aligned}
\label{eq:gyromodel2}
     \mathbf{X}_t & = \sum_{i=1}^{3}\mathbf{b}_{i;t} + \bm{\xi}_t, & \bm{\xi}_t & \simiid \mathcal{N}(\mathbf{0}, \mathbf{R}), \\
     \mathbf{b}_{i;t} & = \phi_{i}\mathbf{b}_{i;t-1} + \bm{\eta}_{i;t}, &  \bm{\eta}_{i;t} & \simiid \mathcal{N}\big(\mathbf{0}, \mathbf{P}_i\big).
\end{aligned}
\end{equation}
While the white noise innovation matrix $\mathbf{R}$ has the same value as in Section~\ref{sec:study_1} with the diagonal elements given in \eqref{eqn:diag_R}, the other parameters are as follows:
\begin{equation*}
    \begin{array}{ccc}
    \phi_1 = 0.9975214, & \phi_2 = 0.9998705, & \phi_3 = 0.9999933.
    \end{array}
\end{equation*}
Moreover, the diagonal elements of $\mathbf{P}_i$, which we denote as $\mathbf{p}_i$, are given by:
\begin{align*}
    \mathbf{p}_1 & = [0.401, 1.178, 1.966, 1.310, 1.928, 1.434] \times 10^{-11}, \nonumber\\
    \mathbf{p}_2 & = [6.714, 5.607, 7.499, 1.208, 7.499, 6.462] \times 10^{-13}, \nonumber\\
    \mathbf{p}_3 & = [5.062, 1.612, 2.103, 0.788, 5.443, 4.990] \times 10^{-14}, \nonumber
\end{align*}
all expressed in $\text{deg}^2/\text{s}^2$. 
Similarly to Section~\ref{sec:study_1}, we present in Figure~\ref{fig:s002_corr} the intensity of the absolute correlation matrices $\mathbf{P}_i$, while the exact values of $\mathbf{P}_i$ are given in Appendix~\ref{app:simparams:case2}.

\begin{figure}[t]
    \centering
    \includegraphics{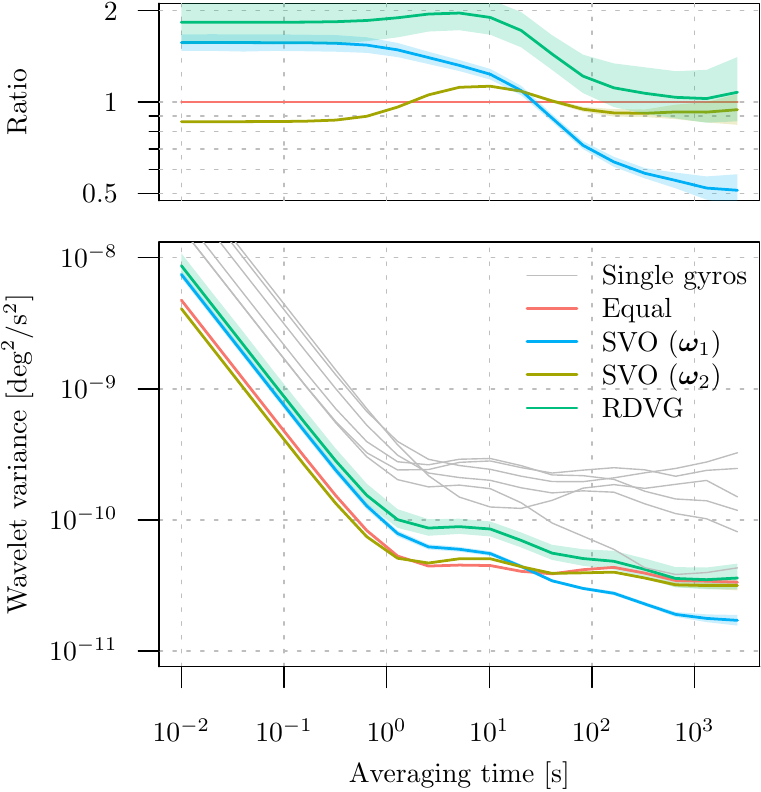}
    \caption{Lower plot: Comparison of the WV of the virtual gyroscopes obtained with different methods considered in the simulation presented in Section~\ref{sec:study_2}. Upper plot: Summary of the ratios of the WV of the virtual gyroscopes based on each method with the WV of the virtual gyroscope computed based on the equally weighted coefficients $\mathbf{c}_{\text{Eq}}$.}
    \label{fig:case2wv}
\end{figure}

The scenario in this simulation is more realistic when dealing with low-cost MEMS IMUs where the WV representations of individual signals do not resemble the typical ``V'' shape of tactical grade IMUs as the ones observed in Figure~\ref{fig:case1wv}, but rather have shapes such as those represented in Figure~\ref{fig:case2wv}. In this case, the stochastic model is much more complex compared to the setting of Section~\ref{sec:study_1}. In practical situations, it would appear difficult to firstly identify the underlying parametric model and successively create an optimal virtual signal for this specific model. Indeed, even assuming that the true model were known, adequately estimating $\mathbf{R}$, $\phi_i$ and $\mathbf{P}_i$ is an additional challenge before finally being able to optimize the corresponding virtual signal. The complicated parametric setting of this simulation is however not an issue for the proposed SVO since, as highlighted previously, it is non-parametric and does not rely on any parametric assumptions on the processes underlying the individual error signals.

\begin{figure}[t]
    \centering
    \includegraphics{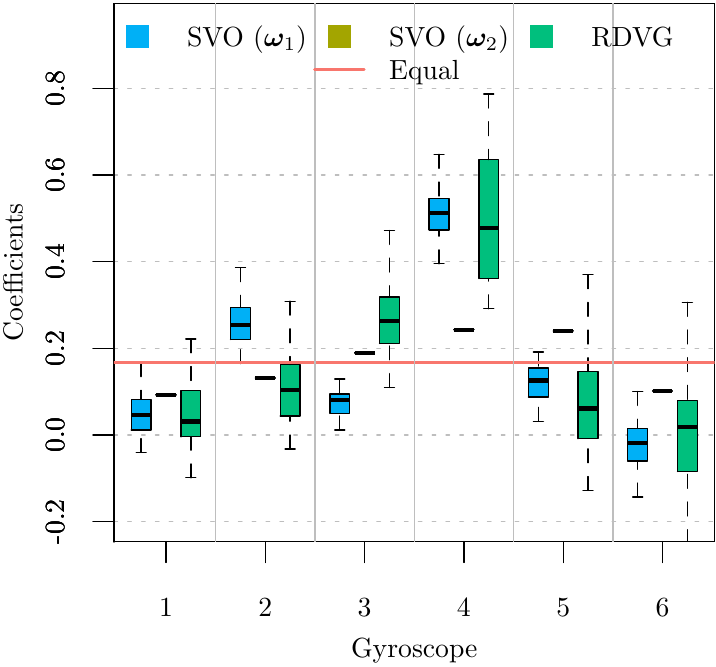}
    \caption{Empirical distributions of the estimated coefficients $\wh{\mathbf{c}}_\text{RDVG}$, $\wh{\mathbf{c}}_{\bomega_1}$ and $\wh{\mathbf{c}}_{\bomega_2}$ computed in the simulation presented in Section~\ref{sec:study_2}.}
    \label{fig:case2coeffs}
\end{figure}

We consider the same setup as in the simulation presented in Section~\ref{sec:study_1}. That is, we consider the same two sets of weights, $\bomega_1$ and $\bomega_2$, respectively focusing on optimizing the small and large levels of WV (i.e. short- and long-scale variance) of the virtual signals, and we use the SVO to estimate the corresponding coefficients $\wh{\mathbf{c}}_{\bomega_1}$ and $\wh{\mathbf{c}}_{\bomega_2}$. The values of both $\bomega_1$ and $\bomega_2$ are provided in Appendix~\ref{app:simparams:case2}. Although the model assumptions are violated, we also consider the method in~\cite{vaccaro2017reduced} to study its performance in this mis-specified parametric setting. As in Section~\ref{sec:study_1}, we consider the equally weighted approach based on $\mathbf{c}_\text{Eq}$ as a benchmark. We present the comparison of the WV of the virtual signals delivered by each method in Figure~\ref{fig:case2wv}. We also compare the empirical distributions of the estimated coefficients for each method in Figure~\ref{fig:case2coeffs}. Both Figures~\ref{fig:case2wv} and \ref{fig:case2coeffs} can be interpreted in the same manner as Figures~\ref{fig:case1wv} and \ref{fig:case1coeffs} in Section~\ref{sec:study_1}.

From Figures~\ref{fig:case2wv} we can clearly observe that the WV of the virtual signals based on the method of \cite{vaccaro2017reduced} are the largest at all scales among all considered approaches. On the contrary, the SVO based on $\bomega_1$ significantly reduces the WV of the constructed virtual signal at the large time-scales, and the SVO based on $\bomega_2$ minimizes the WV at the short time-scales and still performs better than both the equally weighted approach and the method of \cite{vaccaro2017reduced} at the larger time-scales. From Figure~\ref{fig:case2coeffs}, we can observe that the coefficients based on the method of \cite{vaccaro2017reduced} display a large variability. Moreover, we can also see that the long-scale methods (i.e. methods of \cite{vaccaro2017reduced} and the SVO based on $\bomega_1$) appear to disagree over certain individual signals (e.g. the second and third signals), which possibly entails the better performance of the SVO method based on $\bomega_1$ in terms of the resulting WV that we observe in Figure~\ref{fig:case2wv}.

\section{Case Study}
\label{sec:realcase}
 
In this section, we present an applied case where we consider two classes of gyroscopes available in the InvenSense ICM-20689 and Bosch BMI055 low-cost MEMS IMUs. One device of each was installed in Holybro Pixhawk 4, a popular open-hardware autopilot for unmanned aerial vehicles~\cite{pixhawk}. We collected 15 hours of static data at 200 Hz using one Pixhawk 4 board. In order to mimic more complex settings with multiple redundant signals, we split all signals in half and treated the second half as if they were measured jointly with the first half. This allows us to mimic an array of 12 gyroscopes (6 signals for each type of IMU) with an overall maximum level of wavelet decomposition $J = 21$.

\begin{figure}
    \centering
    \includegraphics{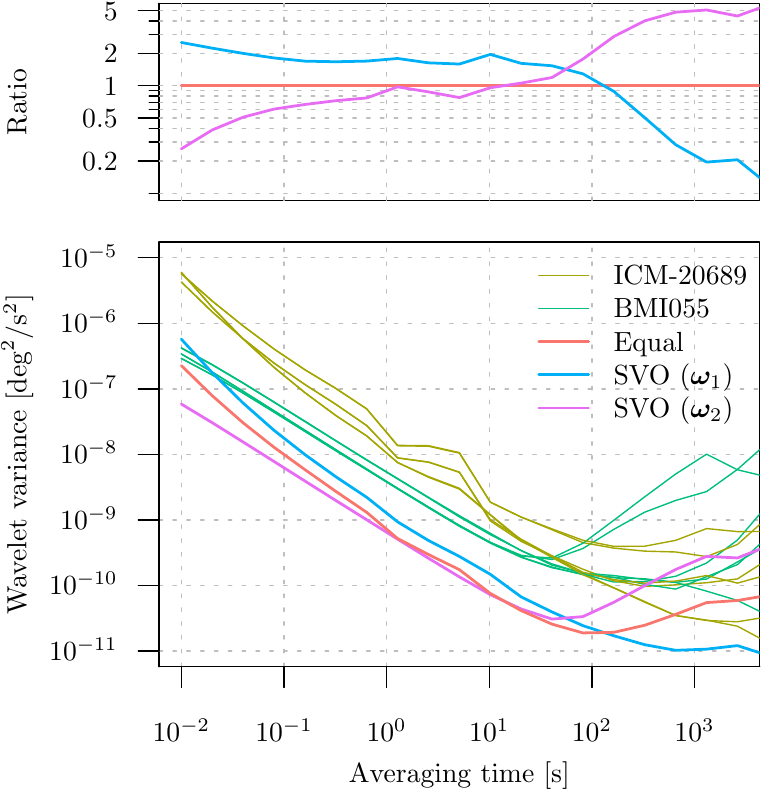}
    \caption{Lower plot: Comparison of the WV of the virtual gyroscopes obtained with different methods considered in the case study in Section~\ref{sec:realcase}. Upper plot: Summary of the ratios of the WV of the virtual gyroscopes based on each method with the WV of the virtual gyroscope computed based on the equally weighted coefficients $\mathbf{c}_{\text{Eq}}$.}
    \label{fig:case3wv}
\end{figure}

The first aspect to underline when observing the WV in Figure~\ref{fig:case3wv} is that the possible identification of a parametric model to optimize the WV of the resulting virtual signal appears quite challenging, especially taking into account the ``bumps'' at the intermediate levels of the ICM-20689. Moreover, the signals from both BMI055 and ICM-20689 are, in some sense, complementary to each other with better properties of the short and long term correlated errors, respectively. {\color{black}This suggests that a combination of the signals from these two devices could deliver a virtual signal that has better properties either at the short- or long-scales, or any combination of the two, depending on the weight vector that the user defines according to the target application.} Similarly to the simulations in Section~\ref{sec:study}, we consider three types of coefficients: one with the equally weighted approach based on $\c_{\text{Eq}}\vcentcolon = 1/12\cdot \1_{12}$, one with the SVO based on $\bomega_1$ ($\c_{\bomega_1}$) that aims at minimizing the variance at the large time-scales, and one with the SVO based on $\bomega_2$ ($\c_{\bomega_2}$) that focuses on the short time-scales. The values of both $\bomega_1$ and $\bomega_2$ are provided in Appendix~\ref{app:case_study}.

\begin{figure}
    \centering
    \includegraphics{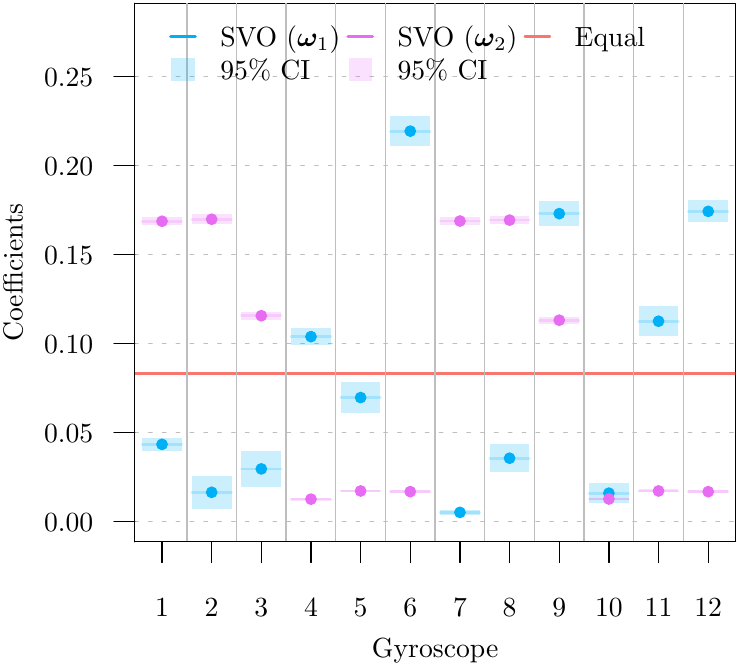}
    \caption{Estimated coefficients using the SVO method based on $\bomega_1$ and $\bomega_2$, with 95\% confidence intervals, considered in the case study in Section~\ref{sec:realcase}.}
    \label{fig:case_coeff}
\end{figure}

As can be observed in Figure~\ref{fig:case3wv}, the SVO based on $\bomega_1$ greatly diminishes the WV of the virtual signal at the large time-scales compared to the equally weighted approach. Moreover, the resulting WV is lower than (or comparable to) the ones of all individual gyroscopes, highlighting a good overall performance for all scales although it focuses mainly on the last ones. On the other hand, the one based on $\bomega_2$ reduces the WV at the short time-scales. In addition, the different shapes of the constructed virtual signals are justified by the different coefficients obtained by each approach as shown in Figure~\ref{fig:case_coeff} along with the 95\% confidence intervals. It can be noticed from Figure~\ref{fig:case_coeff} that, aside from being significantly different from $\c_{\text{Eq}}$, the SVO based on $\bomega_1$ and $\bomega_2$ appears to deliver coefficients that are significantly different from each other (except for the tenth gyroscope), and are often opposite to each other with respect to $\c_{\text{Eq}}$. Based on this plot, it is also possible to identify which individual signals are dominated by a short- or long-scale variance. For example, gyroscopes 1, 2, 3, 7, 8 would appear to have low short-scale variance as the corresponding coefficients in $\wh{\c}_{\bomega_2}$ are higher than the ones in $\wh{\c}_{\bomega_1}$ and $\c_{\text{Eq}}$, whereas the gyroscopes 4, 6, 11, 12 would appear to have low long-scale variance as the corresponding coefficients in $\wh{\c}_{\bomega_1}$ are higher.


\section{Conclusions}
\label{sec:conclusions}

In this paper we propose a new method to combine redundant signals, such as those issued from an array of aligned gyroscopes, to construct an optimal virtual signal with better stochastic properties with respect to the original individual signals. This method is flexible as it allows the users to determine the time-horizon over which the resulting virtual signal should be optimized based on their applications. Statistical properties such as the consistency and asymptotic normality are derived for the coefficients on individual signals. 
Compared to existing alternatives in the literature, the proposed method is non-parametric and thus does not depend on any parametric assumption on the underlying stochastic processes of the individual signals. This provides considerable advantages in real-world applications where, for example, signals representing gyroscope errors can be extremely complex and difficult to model in a parametric manner. As a consequence of our results, we have also verified the properties of a general non-parametric method to adequately estimate the asymptotic covariance matrix of the wavelet cross-covariance (or WCCV) estimator, which is of considerable importance for many other applications beyond the scope of this work. Simulations and case study provide support to the good overall performance of the proposed method, highlighting that this allows to greatly improve the stochastic properties of individual redundant signals and consequently deliver important impacts in the use of sensor arrays for navigation.

\bibliographystyle{unsrt}
\bibliography{ref} 

\appendix

\subsection{Verify Conditions~\ref{cond:stationary} to \ref{cond:stability} for a composite process made of a random walk and a finite sum of AR(1) processes}
\label{proof:verify_cond_rw_ar1}

\begin{proof}
In this section, we consider Haar wavelet filter and therefore, we will verify Conditions~\ref{cond:stationary} to \ref{cond:stability} directly on the difference of the composite process made of a random walk and a finite sum of AR(1) processes. More specifically, we define
\begin{equation*}
\begin{aligned}
    & X_t = X_{t-1} + \alpha A_t, \\
    & Y_t^{(1)} = \phi_1 Y_{t-1}^{(1)} + \beta_1 B_t^{(1)}, \\
    & \ldots \\
    & Y_t^{(m)} = \phi_m Y_{t-1}^{(m)} + \beta_m B_t^{(m)}, \\
    & Z_t = X_t + Y_t^{(1)} + \ldots + Y_t^{(m)},
\end{aligned}   
\end{equation*}
where $\phi_1,\ldots,\phi_m \in (-1,1)$, $\alpha,\beta_1,\ldots,\beta_m$ are finite positive real numbers, $m$ is a finite positive integer, and $A_t, B_t^{(1)}, \ldots, B_t^{(m)} \overset{iid}{\sim} \mathcal{N}(0,1)$. Let $U_t \overset{iid}{\sim} \mathcal{N}(0,1)$, then we have 
\begin{equation*}
\begin{aligned}
    \Delta_t &\vcentcolon = Z_t - Z_{t-1} \\
    &= (X_t-X_{t-1}) + (Y_t^{(1)}-Y_{t-1}^{(1)}) + \ldots + (Y_t^{(m)}-Y_{t-1}^{(m)}) \\
    &= \alpha A_t + \sum_{i=0}^\infty \phi_1^i \beta_1 (B_{t-i}^{(1)}-B_{t-1-i}^{(1)}) + \ldots \\
    &\quad + \sum_{i=0}^\infty \phi_m^i \beta_m (B_{t-i}^{(m)}-B_{t-1-i}^{(m)}) \\
    &= \alpha A_t + \beta_1 B_t^{(1)} + \ldots + \beta_m B_t^{(m)} \\
    &\quad + \sum_{i=1}^\infty \Big\{(\phi_1-1)\phi_1^{i-1}\beta_1B_{t-i}^{(1)} + \ldots \\
    &\quad + (\phi_m-1)\phi_m^{i-1}\beta_mB_{t-i}^{(m)}\Big\} \\
    &\overset{D}{=} (\alpha^2+\beta_1^2+\ldots+\beta_m^2)^{1/2} U_t + \\
    &\quad \sum_{i=1}^\infty \Big\{(\phi_1-1)^2\phi_1^{2(i-1)}\beta_1^2 + \ldots \\
    &\quad + (\phi_m-1)^2\phi_m^{2(i-1)}\beta_m^2\Big\}^{1/2} U_{t-i},
\end{aligned}
\end{equation*}
where the notation $X\overset{D}{=}Y$ represents that $X$ has the same distribution as $Y$. Therefore, Condition~\ref{cond:stationary} is verified. 

Now we consider the verification of Condition~\ref{cond:mom_bound}. We start by deriving the first four moments of $Y_t^{(1)}$. As $Y_t^{(1)} = \sum_{i=0}^\infty \phi_1^i \beta_1 B_{t-i}^{(1)}$, we directly obtain $\mathbb{E}(Y_t^{(1)}) = 0$. Moreover, we have
\begin{equation*}
\begin{aligned}
    \mathbb{E}(Y_t^{(1)\,2}) &= \mathbb{E}(\phi_1^2Y_{t-1}^{(1)\,2} + 2\phi_1\beta_1 Y_{t-1}^{(1)} B_t^{(1)} + \beta_1^2 B_t^{(1)\,2}) \\
    &= \phi_1^2 \mathbb{E}(Y_t^{(1)\,2}) +\beta_1^2,
\end{aligned}
\end{equation*}
which gives $\mathbb{E}(Y_t^{(1)\,2}) = \beta_1^2 / (1-\phi_1^2) <\infty$. For the third moment of $Y_t$, we write
\begin{equation*}
\begin{aligned}
    \mathbb{E}(Y_t^{(1)\,3}) &= \mathbb{E}(\phi_1^3Y_{t-1}^{(1)\,3} + 3\phi_1^2Y_{t-1}^{(1)\,2}\beta_1 B_t^{(1)} \\
    &\quad + 3 \phi_1 Y_{t-1}^{(1)} \beta_1^2 B_t^{(1)\,2} + \beta_1^3 B_t^{(1)\,3}) \\
    &= \phi_1^3 \mathbb{E}(Y_t^{(1)\,3}),
\end{aligned}
\end{equation*}
so $\mathbb{E}(Y_t^{(1)\,3}) = 0$. Lastly, for the fourth moment of $Y_t^{(1)}$, we write
\begin{equation*}
\begin{aligned}
    \mathbb{E}(Y_t^{(1)\,4}) &= \mathbb{E}(\phi_1^4Y_{t-1}^{(1)\,4} + 4\phi_1^3Y_{t-1}^{(1)\,3}\beta_1 B_t^{(1)} \\
    &\quad + 6\phi_1^2Y_{t-1}^{(1)\,2}\beta_1^2 B_t^{(1)\,2} + 4\phi_1 Y_{t-1}^{(1)}\beta_1^3 B_t^{(1)\,3} \\
    &\quad + \beta_1^4 B_t^{(1)\,4}) \\
    &= \phi_1^4 \mathbb{E}(Y_t^{(1)\,4}) + 6\phi_1^2\beta_1^2 \mathbb{E}(Y_t^{(1)\,2}) + 3\beta_1^4 \\
    &= \phi_1^4 \mathbb{E}(Y_t^{(1)\,4}) + \frac{6\phi_1^2\beta_1^4}{1-\phi_1^2} + 3\beta_1^4,
\end{aligned}
\end{equation*}
and thus,
\begin{equation*}
    \mathbb{E}(Y_t^{(1)\,4}) = \frac{6\phi_1^2\beta_1^4}{(1-\phi_1^2)(1-\phi_1^4)} + \frac{3\beta_1^4}{1-\phi_1^4} < \infty.
\end{equation*}
Similarly, we can show that $\mathbb{E}(Y_t^{(i)\,j})<\infty$ for $i=2,\ldots,m$ and $j=1,\ldots,4$. Since the first four moments of $A_t, Y_t^{(1)},\ldots, Y_t^{(m)}$ are finite, we have
\begin{equation*}
\begin{aligned}
    \mathbb{E}(\Delta_t^4) &= \mathbb{E}\bigg[\Big\{(X_t-X_{t-1}) + (Y_t^{(1)}-Y_{t-1}^{(1)}) + \ldots \\
    &\quad + (Y_t^{(m)}-Y_{t-1}^{(m)})\Big\}^4\bigg] \\
    &= \mathbb{E}\bigg[\Big\{\alpha A_t + (Y_t^{(1)}-Y_{t-1}^{(1)}) + \ldots \\
    &\quad + (Y_t^{(m)}-Y_{t-1}^{(m)})\Big\}^4\bigg]\\
    &<\infty,
\end{aligned}
\end{equation*}
which verifies Condition \ref{cond:mom_bound}. 

We remain to verify Condition \ref{cond:stability}. If $i=0$, we define $$\delta_i \vcentcolon=(\alpha^2+\beta_1^2+\ldots+\beta_m^2)^{1/2},$$ and if $i \geq 1$, we define $$\delta_i \vcentcolon= \left[(\phi_1-1)^2\phi_1^{2(i-1)}\beta_1^2 + \ldots + (\phi_m-1)^2\phi_m^{2(i-1)}\beta_m^2\right]^{1/2}.$$
Since $\phi_1,\ldots,\phi_m,\alpha,\beta_1,\ldots,\beta_m$ are all finite real numbers and $m$ is also finite, we have $\delta_i$ to be finite for all $i \geq 0$. Then if $t=0$, we define $$\Delta_t^* \vcentcolon= \delta_0 U_0^* + \sum_{i=1}^\infty \delta_i U_{-i},$$ and if $t>0$, we define $$\Delta_t^* \vcentcolon= \delta_0 U_t + \sum_{i=1}^{t-1} \delta_i U_{t-i} + \delta_t U_0^* + \sum_{i=t+1}^\infty \delta_i U_{t-i},$$
where $U_0^*$ is also i.i.d. $\mathcal{N}(0,1)$ random variable. So we have $\Delta_t - \Delta_t^* = \delta_t (U_0 - U_0^*)$. Note that
\begin{equation*}
\begin{aligned}
    \sum_{t=1}^\infty \delta_t &= \sum_{t=1}^\infty \left(\sum_{j=1}^m(\phi_j-1)^2\phi_j^{2(t-1)}\beta_j^2\right)^{1/2} \\
    &\leq \sum_{t=1}^\infty \sum_{j=1}^m (1-\phi_j) |\phi_j|^{t-1} \beta_j \\
    &= \sum_{j=1}^m (1-\phi_j)\beta_j \sum_{t=1}^\infty |\phi_j|^{t-1} \\
    &= \sum_{j=1}^m \frac{(1-\phi_j)\beta_j}{1-|\phi_j|} \\
    &< \infty.
\end{aligned}
\end{equation*}
So we have
\begin{equation*}
\begin{aligned}
    &\sum_{t=0}^\infty \mathbb{E}[(\Delta_t - \Delta_t^*)^4]^{1/4} = \mathbb{E}[(U_0-U_0^*)^4]^{1/4} \sum_{t=0}^\infty \delta_t \\
    &\quad = \mathbb{E}[(U_0-U_0^*)^4]^{1/4}\delta_0 + \mathbb{E}[(U_0-U_0^*)^4]^{1/4}\sum_{t=1}^\infty \delta_t <\infty,
\end{aligned}
\end{equation*}
where the last inequality is because the first four moments of both $U_0$ and $U_0^*$ are finite, $\delta_0$ is finite and $\sum_{t=1}^\infty \delta_t$ is finite. Therefore, Condition \ref{cond:stability} is verified. 
\end{proof}

\subsection{Proof of Theorem~\ref{thm:block_boot}}
\label{proof:block_boot}

\begin{proof}
To prove the statement in Theorem~\ref{thm:block_boot}, we essentially want to show that
\begin{equation*}
    \sqrt{T}\wh{\bm{\Sigma}}^{*\, -1/2} (\wh{\c} - \c_0) \overset{D}{\to} \mathcal{N}(\0, \mathbf{I}_p),
\end{equation*}
where the left hand side can also be expressed as 
\begin{equation}
\label{eqn:pf_block_boot_5}
    \sqrt{T}\bm{\Sigma}^{-1/2} (\wh{\c} - \c_0) + \sqrt{T}\left(\wh{\bm{\Sigma}}^{*\, -1/2} -  \bm{\Sigma}^{-1/2} \right) (\wh{\c} - \c_0).
\end{equation}
By Theorem~\ref{thm:asymp_norm}, we have $\sqrt{T}\bm{\Sigma}^{-1/2} (\wh{\c} - \c_0) \overset{D}{\to} \mathcal{N}(\bm{0}, \mathbf{I}_p)$, implying that
\begin{equation}
\label{eqn:pf_block_boot_6}
    \lim_{T \to \infty}\,  \sup_{z \in \real} \; \left|\Pr\left[\sqrt{T}\Sigma_{i,i}^{-1/2} (\wh{c}_i - c_{0,i}) \leq z\right] - \Phi(z)\right| = 0,
\end{equation}
where $\Phi(\cdot)$ denotes the cumulative distribution function of the standard normal distribution (see e.g. Lemma 2.11 in \cite{van2000asymptotic}). Consequently, we obtain
\begin{equation*}
    \lim_{T \to \infty}\,  \Pr\left[c_{0,i} \in \left(\wh{c}_i \pm z_{1-\alpha/2} \sqrt{\Sigma_{i,i}/T}\right)\right] = 1-\alpha,
\end{equation*}
and therefore, our goal is to determine the asymptotic behavior of $\sqrt{T}\left(\wh{\bm{\Sigma}}^{*\, -1/2} -  \bm{\Sigma}^{-1/2} \right) (\wh{\c} - \c_0)$.  Since $\wh{\bm{\Sigma}}^{*} = \wh{\G} \wh{\V}^* \wh{\G}\trans$, we want to determine the asymptotic behaviors of $\wh{\V}^*$ and $\wh{\G}$.

We first focus on the matrix $\wh{\V}^*$ issued from the MBB procedure. Let us define $\wt{\V}$ as the asymptotic covariance matrix of $\wh{\bg}^{(s)} = \{\wh{\gamma}_{i,k,j}^{(s)}\}_{\substack{j=1,\ldots,J \\ i,k=1,\ldots,p}}$ with 
\begin{equation*}
    \wh{\gamma}_{i,k,j}^{(s)} \vcentcolon = \frac{1}{M_J} \sum_{t=1}^{M_J} W_{i,j,t} W_{k,j,t}.
\end{equation*}
Similarly, we define $\wh{\bg}^{(s)*}=\{\wh{\gamma}_{i,k,j}^{(s)*}\}_{\substack{j=1,\ldots,J \\ i,k=1,\ldots,p}}$ with
\begin{equation*}
    \wh{\gamma}_{i,k,j}^{(s)*} \vcentcolon = \frac{1}{M_J} \sum_{t=1}^{M_J} W_{i,j,t}^* W_{k,j,t}^*,
\end{equation*}
based on which we define $\wt{\V}^*$ below as an estimator for $\wt{\V}$:
\begin{equation*}
    \wt{\V}^* \vcentcolon= \frac{T}{H} \sum_{h=1}^H \left\{\wh{\bg}_h^{(s)*} - \wh{\bg}^{(s)}\right\}\left\{\wh{\bg}_h^{(s)*} - \wh{\bg}^{(s)}\right\}\trans.
\end{equation*}
Under Conditions \ref{cond:stationary} to \ref{cond:stability}, by the results in \cite{buhlmann2002bootstraps} and Cauchy-Schwarz inequality, we have 
\begin{equation*}
    \mathbb{E}\left[ (\wt{V}_{i,k}^* - \wt{V}_{i,k})^2\right] = \mathcal{O}(T^{-2/3}).
\end{equation*}
So for sufficiently large $T$ we have 
\begin{equation*}
\begin{aligned}
    \Pr\left[\bigg|T^{1/3}\left(\wt{V}_{i,k}^* - \wt{V}_{i,k}\right)\bigg| > m\right] &<  \frac{\mathbb{E}\left[T^{2/3} (\wt{V}_{i,k}^* - \wt{V}_{i,k})^2\right]}{m^2} \\
    &< C,
\end{aligned}
\end{equation*}
where $C$ is some finite positive constant. Here the first inequality is by Markov's inequality, and the last inequality is because $\mathbb{E}[ (\wt{V}_{i,k}^* - \wt{V}_{i,k})^2] = \mathcal{O}(T^{-2/3})$. Therefore, for all $i, k = 1,\ldots, Jp^2$ we have
\begin{equation}
\label{eqn:pf_block_boot_4}
    \wt{V}_{i,k}^* = \wt{V}_{i,k} + \mathcal{O}_{\rm p}(T^{-1/3}).
\end{equation}
Next we want to check the asymptotic difference between $\wh{\bg}$ (the WCCV estimator computed on the original untrimmed wavelet coefficients) and $\wh{\bg}^{(s)}$ (the WCCV estimator computed on the trimmed wavelet coefficients). Elementwise we can observe the following difference:
\begin{equation}
\begin{aligned}
\label{eqn:pf_block_boot_1}
    \wh{\gamma}_{i,k,j} - \wh{\gamma}_{i,k,j}^{(s)} &=  \frac{1}{M_j} \sum_{t=1}^{M_j} W_{i,j,t} W_{k,j,t} - \frac{1}{M_J} \sum_{t=1}^{M_J} W_{i,j,t} W_{k,j,t} \\
    &= \frac{1}{M_j} \left( \sum_{t=1}^{M_J}W_{i,j,t} W_{k,j,t} + \sum_{t=M_J+1}^{M_j} W_{i,j,t} W_{k,j,t} \right) \\
    &\quad - \frac{1}{M_J} \sum_{t=1}^{M_J} W_{i,j,t} W_{k,j,t} \\
    &= -\frac{D_j}{M_JM_j} \sum_{t=1}^{M_J} W_{i,j,t} W_{k,j,t} \\
    &\quad + \frac{1}{M_j}\sum_{t=M_J+1}^{M_j} W_{i,j,t} W_{k,j,t}.
\end{aligned}
\end{equation}
Under Condition \ref{cond:stationary}, we have that 
\begin{equation}
\begin{aligned}
\label{eqn:pf_block_boot_2}
    &-\frac{D_j}{M_JM_j} \sum_{t=1}^{M_J} W_{i,j,t} W_{k,j,t} \\
    &= -\frac{D_j}{M_j\sqrt{M_J}} \left[\frac{1}{\sqrt{M_J}}\sum_{t=1}^{M_J} (W_{i,j,t} W_{k,j,t} - \gamma_{i,k,j})\right] - \frac{D_j}{M_j}\gamma_{i,k,j} \\
    &= -\frac{D_j}{M_j\sqrt{M_J}} \mathcal{O}_{\rm p}(1) - \frac{D_j}{M_j} \gamma_{i,k,j} = \mathcal{O}_{\rm p}(T^{-1}),
\end{aligned}
\end{equation}
where the second equality is by the result of \cite{xu2019multivariate}, and
\begin{equation}
\label{eqn:pf_block_boot_3}
    \frac{1}{M_j}\sum_{t=M_J+1}^{M_j} W_{i,j,t} W_{k,j,t} = \mathcal{O}_{\rm p}\left(T^{-1}\right).
\end{equation}
Putting the results of \eqref{eqn:pf_block_boot_2} and \eqref{eqn:pf_block_boot_3} into \eqref{eqn:pf_block_boot_1}, we have elementwise that
\begin{equation*}
    \wh{\bg} - \wh{\bg}^{(s)} = \mathcal{O}_{\rm p}(T^{-1}).
\end{equation*}
Therefore, we have
\begin{equation*}
\begin{aligned}
    \sqrt{T}\left\{\wh{\bg}^{(s)} - \bg_0\right\} &= \sqrt{T}\left\{\wh{\bg}^{(s)} - \wh{\bg}\right\} + \sqrt{T}\left(\wh{\bg} - \bg_0\right) \\
    &= \mathcal{O}_{\rm p}(T^{-1/2}) + \sqrt{T}\left(\wh{\bg} - \bg_0\right) \overset{D}{\to} \mathcal{N}(\0, \V),
\end{aligned}
\end{equation*}
which implies that the asymptotic distribution of $\wh{\bg}^{(s)}$ is the same as the one of $\wh{\bg}$, and thus, $\wt{\V} = \V$. Using the same arguments above we can also obtain that $\wh{\bg}^* - \wh{\bg}^{(s)*} = \mathcal{O}_{\rm p}(T^{-1})$ and $\wt{\V}^* = \wh{\V}^*$. So we obtain
\begin{equation*}
    \wh{V}^*_{i,k} = \wt{V}^*_{i,k} = \wt{V}_{i,k} + \mathcal{O}_{\rm p}(T^{-1/3}) = V_{i,k} + \mathcal{O}_{\rm p}(T^{-1/3}),
\end{equation*}
where the second equality is by \eqref{eqn:pf_block_boot_4}. In addition, by applying the multivariate delta method we can obtain $\wh{G}_{i,k} = G_{i,k} + \mathcal{O}_{\rm p}(T^{-1/2})$. Therefore, we have
\begin{equation*}
\begin{aligned}
    \wh{\Sigma}_{i,k}^* &= \sum_{j=1}^{Jp^2}\sum_{l=1}^{Jp^2} \wh{G}_{i,j} \wh{V}_{j,l}^{*} \wh{G}_{l,k}\trans \\
    &= \sum_{j=1}^{Jp^2}\sum_{l=1}^{Jp^2} \left\{G_{i,j}+\mathcal{O}_{\rm p}(T^{-1/2})\right\} \left\{V_{j,l}+\mathcal{O}_{\rm p}(T^{-1/3})\right\} \\
    &\quad \left\{G_{l,k}\trans + \mathcal{O}_{\rm p}(T^{-1/2})\right\} \\
    &= \sum_{j=1}^{Jp^2}\sum_{l=1}^{Jp^2} \left\{G_{i,j}V_{j,l}G_{l,k}\trans + \mathcal{O}_{\rm p}(T^{-1/6})\right\} \\
    &\quad = \Sigma_{i,k}+ \mathcal{O}_{\rm p}(T^{-1/6}),
\end{aligned}
\end{equation*}
which implies $(\wh{\Sigma}_{i,k}^{*\, -1/2} - \Sigma_{i,k}^{-1/2} ) = \mathcal{O}_{\rm p}(T^{-1/6})$ by delta method. Thus, with the asymptotic normality result of $\wh{\c}$ from Theorem~\ref{thm:asymp_norm} that $\sqrt{T}(\wh{\c}-\c_0) = \mathcal{O}_{\rm p}(1)$, we obtain elementwise that 
\begin{equation*}
    \sqrt{T}\left(\wh{\bm{\Sigma}}^{*\, -1/2} -  \bm{\Sigma}^{-1/2} \right) (\wh{\c} - \c_0) = \mathcal{O}_{\rm p}(T^{-1/6}) = o_{\rm p}(1),
\end{equation*}
and thus, using the relation of \eqref{eqn:pf_block_boot_5} and the result of \eqref{eqn:pf_block_boot_6}, we obtain
\begin{equation*}
\begin{aligned}
    &\lim_{T \to \infty}\,  \Pr\left[c_{0,i} \in \left(\wh{c}_i \pm z_{1-\alpha/2} \sqrt{\wh{\Sigma}^*_{i,i}/T}\right)\right] \\
    &= \lim_{T \to \infty}\,  \Pr\left[c_{0,i} \in \left(\wh{c}_i \pm z_{1-\alpha/2} \sqrt{\Sigma_{i,i}/T}\right)\right] = 1-\alpha,
\end{aligned}
\end{equation*}
which completes the proof.
\end{proof}

\subsection{Settings for simulation studies in Section~\ref{sec:study}}
\subsubsection{Case 1}
\label{app:simparams:case1}

In this section, we provide the parameter values that we used in the simulation presented in Section~\ref{sec:study_1}. 

The matrices $\mathbf{R}$ and $\mathbf{Q}$, as introduced in~\eqref{eq:gyromodel1}, are given as follows:
\begin{equation*}
\mathbf{R} = \left[
\resizebox{8cm}{!}{
$\begin{array}{r r r r r r} 
 2.805556 & 0.000000 & 0.000000 & 0.000000 & 0.000000 & 0.000000 \\
0.000000 & 1.977778 & 0.000000 & 0.000000 & 0.000000 & 0.000000 \\
0.000000 & 0.000000 & 1.361111 & 0.000000 & 0.000000 & 0.000000 \\
0.000000 & 0.000000 & 0.000000 & 1.063889 & 0.000000 & 0.000000 \\
0.000000 & 0.000000 & 0.000000 & 0.000000 & 1.069444 & 0.000000 \\
0.000000 & 0.000000 & 0.000000 & 0.000000 & 0.000000 & 2.538889 \\
\end{array}$}
\right],
\end{equation*}
with the unit $\text{deg}^2/\text{s}^2 \times 10^{7}$, and
\begin{equation*}
\mathbf{Q} = \left[
\resizebox{8cm}{!}{
$\begin{array}{r r r r r r} 
 0.255058 & -0.008573 & 0.102881 & -0.162894 & -0.240055 & -0.055727 \\
-0.008573 & 0.471536 & 0.199331 & 0.010717 & 0.032150 & 0.207905 \\
0.102881 & 0.199331 & 3.489369 & -1.281722 & 0.055727 & -0.302212 \\
-0.162894 & 0.010717 & -1.281722 & 2.091907 & 0.409379 & 0.544410 \\
-0.240055 & 0.032150 & 0.055727 & 0.409379 & 0.555127 & 0.244342 \\
-0.055727 & 0.207905 & -0.302212 & 0.544410 & 0.244342 & 2.501286 \\
\end{array}$}
\right],
\end{equation*}
with the unit $\text{deg}^2/\text{s}^2 \times 10^{13}$.

The weight vectors $\bomega_1$ and $\bomega_2$ are given as follows: 
\begin{align*}
\bomega_1 = [ & \begin{array}{rrrrrrrr}
    0.000, & 0.000, & 0.000, & 0.000, & 0.000, & 0.000,
    \end{array} \\ 
& \begin{array}{rrrrrrrr}
    0.000, & 0.001, & 0.003, & 0.007, & 0.018, & 0.041,
    \end{array} \\ 
& \begin{array}{rrrrrrrr}
    0.077, & 0.112, & 0.135, & 0.147, & 0.151, & 0.153,
    \end{array} \\ 
& \begin{array}{rrrrrrrr}
    0.153
    \end{array}], \\
\bomega_2 = [& \begin{array}{rrrrrrrr}
    0.118, & 0.118, & 0.117, & 0.117, & 0.116, & 0.112,
    \end{array} \\ 
& \begin{array}{rrrrrrrr}
    0.104, & 0.086, & 0.059, & 0.032, & 0.014, & 0.006,
    \end{array} \\ 
& \begin{array}{rrrrrrrr}
    0.002, & 0.001, & 0.000, & 0.000, & 0.000, & 0.000,
    \end{array} \\ 
& \begin{array}{rrrrrrrr}
    0.000
    \end{array} ].
\end{align*}

\subsubsection{Case 2}
\label{app:simparams:case2}

In this section, we provide the parameter values that we used in the simulation presented in Section~\ref{sec:study_2}.

The matrix $\mathbf{R}$ is the same as in Appendix~\ref{app:simparams:case1}, i.e.
\begin{equation*}
\mathbf{R} = \left[
\resizebox{8cm}{!}{
$\begin{array}{r r r r r r} 
 2.805556 & 0.000000 & 0.000000 & 0.000000 & 0.000000 & 0.000000 \\
0.000000 & 1.977778 & 0.000000 & 0.000000 & 0.000000 & 0.000000 \\
0.000000 & 0.000000 & 1.361111 & 0.000000 & 0.000000 & 0.000000 \\
0.000000 & 0.000000 & 0.000000 & 1.063889 & 0.000000 & 0.000000 \\
0.000000 & 0.000000 & 0.000000 & 0.000000 & 1.069444 & 0.000000 \\
0.000000 & 0.000000 & 0.000000 & 0.000000 & 0.000000 & 2.538889 \\
\end{array}$}
\right],
\end{equation*}
with the unit $\text{deg}^2/\text{s}^2 \times 10^{7}$. The matrices $\mathbf{P}_i$ with $i =1,2,3$, as introduced in~\eqref{eq:gyromodel2}, are given as follows:
\begin{equation*}
\mathbf{P}_1 = \left[
\resizebox{8cm}{!}{
$\begin{array}{r r r r r r} 
0.400813 & -0.018061 & -0.044770 & -0.104857 & 0.254671 & -0.234950 \\
-0.018061 & 1.177847 & -0.012905 & -0.374198 & -0.051101 & 0.284942 \\
-0.044770 & -0.012905 & 1.965775 & 0.202629 & -0.175089 & 0.144629 \\
-0.104857 & -0.374198 & 0.202629 & 1.309887 & 0.023176 & 0.412761 \\
0.254671 & -0.051101 & -0.175089 & 0.023176 & 1.928375 & 0.571453 \\
-0.234950 & 0.284942 & 0.144629 & 0.412761 & 0.571453 & 1.434466 \\
\end{array}$}
\right],
\end{equation*}
with the unit $\text{deg}^2/\text{s}^2 \times 10^{11}$,
\begin{equation*}
\mathbf{P}_2 = \left[
\resizebox{8cm}{!}{
$\begin{array}{r r r r r r} 
6.713833 & 1.515552 & 1.245013 & -0.071179 & 0.428489 & 0.957757 \\
1.515552 & 5.607230 & 2.264650 & -0.773724 & 0.281054 & 1.440987 \\
1.245013 & 2.264650 & 7.499051 & -0.162687 & -0.529317 & -0.999575 \\
-0.071179 & -0.773724 & -0.162687 & 1.207920 & -0.187305 & 0.681702 \\
0.428489 & 0.281054 & -0.529317 & -0.187305 & 7.499327 & 2.414762 \\
0.957757 & 1.440987 & -0.999575 & 0.681702 & 2.414762 & 6.461646 \\
\end{array}$}
\right],
\end{equation*}
with the unit $\text{deg}^2/\text{s}^2 \times 10^{13}$, and
\begin{equation*}
\mathbf{P}_3 = \left[
\resizebox{8cm}{!}{
$\begin{array}{r r r r r r} 
5.062006 & -0.643280 & 0.326013 & 0.697800 & 1.727949 & 0.738584 \\
-0.643280 & 1.612038 & -0.318449 & -0.383634 & -0.768214 & -0.100044 \\
0.326013 & -0.318449 & 2.102923 & 0.091458 & 0.398505 & -0.137760 \\
0.697800 & -0.383634 & 0.091458 & 0.788177 & -0.129179 & 0.519119 \\
1.727949 & -0.768214 & 0.398505 & -0.129179 & 5.442670 & 0.378632 \\
0.738584 & -0.100044 & -0.137760 & 0.519119 & 0.378632 & 4.990004 \\
\end{array}$}
\right],
\end{equation*}
with the unit $\text{deg}^2/\text{s}^2 \times 10^{14}$.

The weight vectors $\bomega_1$ and $\bomega_2$ are the same as in Appendix~\ref{app:simparams:case1}, i.e.
\begin{align*}
\bomega_1 = [ & \begin{array}{rrrrrrrr}
    0.000, & 0.000, & 0.000, & 0.000, & 0.000, & 0.000,
    \end{array} \\ 
& \begin{array}{rrrrrrrr}
    0.000, & 0.001, & 0.003, & 0.007, & 0.018, & 0.041,
    \end{array} \\ 
& \begin{array}{rrrrrrrr}
    0.077, & 0.112, & 0.135, & 0.147, & 0.151, & 0.153,
    \end{array} \\ 
& \begin{array}{rrrrrrrr}
    0.153
    \end{array}], \\
\bomega_2 = [& \begin{array}{rrrrrrrr}
    0.118, & 0.118, & 0.117, & 0.117, & 0.116, & 0.112,
    \end{array} \\ 
& \begin{array}{rrrrrrrr}
    0.104, & 0.086, & 0.059, & 0.032, & 0.014, & 0.006,
    \end{array} \\ 
& \begin{array}{rrrrrrrr}
    0.002, & 0.001, & 0.000, & 0.000, & 0.000, & 0.000,
    \end{array} \\ 
& \begin{array}{rrrrrrrr}
    0.000
    \end{array} ].
\end{align*}

\subsection{Setting for case study in Section~\ref{sec:realcase}}
\label{app:case_study}

In this section, we provide the parameter values that we used in the case study presented in Section~\ref{sec:realcase}. In particular, the weight vectors $\bomega_1$ and $\bomega_2$ are given as follows: 
\begin{align*}
\bomega_1 = [ & \begin{array}{rrrrrrrr}
    0.000, & 0.000, & 0.000, & 0.000, & 0.000, & 0.000,
    \end{array} \\ 
& \begin{array}{rrrrrrrr}
    0.000, & 0.000, & 0.000, & 0.001, & 0.003, & 0.007,
    \end{array} \\ 
& \begin{array}{rrrrrrrr}
    0.018, & 0.041, & 0.077, & 0.112, & 0.135, & 0.147,
    \end{array} \\ 
& \begin{array}{rrrrrrrr}
    0.151, & 0.153, & 0.153
    \end{array} ], \\
\bomega_2 = [& \begin{array}{rrrrrrrr}
    0.105, & 0.105, & 0.105, & 0.105, & 0.105, & 0.103,
    \end{array} \\ 
& \begin{array}{rrrrrrrr}
    0.100, & 0.093, & 0.077, & 0.053, & 0.028, & 0.013,
    \end{array} \\ 
& \begin{array}{rrrrrrrr}
    0.005, & 0.002, & 0.001, & 0.000, & 0.000, & 0.000,
    \end{array} \\ 
& \begin{array}{rrrrrrrr}
    0.000, & 0.000, & 0.000
    \end{array} ].
\end{align*}

\end{document}